%% file: main.tex
\documentclass[a4paper,11pt]{article}

\input{MathPhysDef}

\providecommand{\Tr}{\operatorname{Tr}}

\providecommand{\overline}{\overline}

\newcommand{\Ups}{\Upsilon}

\makeatletter
\newsavebox{\@brx}
\newcommand{\llangle}[1][]{\savebox{\@brx}{\(\m@th{#1\langle}\)}%
  \mathopen{\copy\@brx\kern-0.5\wd\@brx\usebox{\@brx}}}
\newcommand{\rrangle}[1][]{\savebox{\@brx}{\(\m@th{#1\rangle}\)}%
  \mathclose{\copy\@brx\kern-0.5\wd\@brx\usebox{\@brx}}}
\makeatother

\usepackage{fontawesome5} 

\usepackage[misc]{ifsym} 

\usepackage{tikz-feynman}

\usetikzlibrary{decorations.markings}

\usetikzlibrary{decorations.pathmorphing,decorations.markings}

\usetikzlibrary{patterns}

\allowdisplaybreaks

\usetikzlibrary{decorations.markings}

\tikzset{
    partial ellipse/.style args={#1:#2:#3}{
        insert path={+ (#1:#3) arc (#1:#2:#3)}
    }
}

\tikzset{->-/.style={decoration={
  markings,
  mark=at position #1 with {\arrow{>}}},postaction={decorate}}}
\tikzset{-<-/.style={decoration={
  markings,
  mark=at position #1 with {\arrow{<}}},postaction={decorate}}}

\begin{document}

\flushbottom

\title{Quantum Monge-Kantorovich optimal transport in spacetime through spatiotemporal Kirkwood-Dirac distributions and spatiotemporal states}

\author[a,\star]{Zhian Jia\orcidlink{0000-0001-8588-173X}}

\affiliation[a]{Institute of Quantum Physics, School of Physics, Central South University,
Changsha 410012, China}

\affiliation[\star]{Email: \href{mailto:giannjia@foxmail.com}{giannjia@foxmail.com}}

\abstract{Monge-Kantorovich optimal transport is a mathematical framework that determines the most efficient way to transform one probability distribution into another while minimizing a given cost function. Its central constructions include the Wasserstein distances, which quantify the space of probability distributions, and dual formulations that link to convex analysis.
Its standard quantum generalization replaces probability distributions with density operators and cost function with positive semidefinite cost operator, which should be regarded as spatial quantum optimal transport. In this work, we propose the quantum  optimal transport problem in spacetime through the spatiotemporal phase-space Kirkwood-Dirac quasiprobability distribution and spatiotemporal states and through spatiotemporal state. We show that the two formalisms are equivalent and related via spatiotemporal operator frames. In the Kirkwood-Dirac phase-space formalism, the spatiotemporal distribution and the cost function are complex-valued, and the spatiotemporal couplings are  Kirkwood-Dirac  quasiprobability distributions that yield the correct marginals for the local states. In the spatiotemporal state formalism, the cost operators are generally non-Hermitian, and the spatiotemporal couplings are generally non-Hermitian and yield the correct reduced local states. The  Margenau-Hill phase-space distribution gives rise to a real-valued optimal transport in spatiotemporal phase space; in the spatiotemporal state formalism, this corresponds to a Hermitian spatiotemporal coupling state. We give a rigorous definitions and a careful discussion of its duality. We also introduce a spatiotemporal quantum Wasserstein metric based on a metric in spatiotemporal phase spaces and show that it satisfies the metric axioms.
}


\maketitle

\section{Introduction}

The asymmetry between space and time gives rise to conceptual difficulties known as the problem of time when one attempts to reconcile quantum mechanics with relativity, and it becomes especially significant in quantum field theory and quantum gravity~\cite{Page1983evolution,Rovelli1991time}. One approach for this is the spatiotemporal quantum mechanics, where a local Hilbert space is assigned to each spacetime point rather than just consider the vector bundle over spatial manifold, namely $\cH_{\rm tot}=\bigotimes_{(x,t)}\cH_{x,t}$,  and the notion of a spatial joint density operator is extended to that of a spatiotemporal state.  Examples include consistent histories~\cite{griffiths1984consistent}, the Feynman-Vernon influence functional~\cite{Feynman1963influenceMatrix} and the influence matrix~\cite{Lerose2021influencematrix}, multiple-time states~\cite{Watanabe1955twostateQM,aharonov2007twostatevectorformalismqauntum,Aharonov2009multi}, pseudo-density operators~\cite{fitzsimons2015quantum} (equivalently, states over time~\cite{fullwood2022quantum,Parzygnat2023pdo}),  quantum combs~\cite{Chiribella2009comb} (equivalently quantum-classical games~\cite{gutoski2007toward}, process tensors~\cite{Pollock2018processtensor}), process matrices~\cite{oreshkov2012quantum} which can capture the indefinite causal order, superdensity operators~\cite{cotler2018superdensity}, doubled density operators and spatiotemporal states~\cite{jia2024spatiotemporal,Jia2025TemporalKirkwoodDirac,jia2026temporaltomography,jia2026unifiedspatiotemporalquantumstates}, and spacetime quantum mechanics~\cite{diaz2024STQM,diaz2026unifyingspacetimeapproachesquantum,diaz2024STQMBosonFermi}, also see \cite{Guo2026Spacetime,Lie2025stateovertime,milekhin2025observable,Glorioso2024spacetimeMutual}. Clarifying how these approaches connect to and differ from one another remains an important open direction~\cite{liu2023unification,Parzygnat2023pdo,fullwood2022quantum,Jia2025TemporalKirkwoodDirac,diaz2026unifyingspacetimeapproachesquantum,jia2026unifiedspatiotemporalquantumstates}.
In \cite{Jia2025TemporalKirkwoodDirac,jia2026unifiedspatiotemporalquantumstates}, we use spatiotemporal Kirkwood-Dirac phase-space distribution to give a unification of some existing frameworks. This unification provides a natural language in which to formulate quantum transport directly in spacetime.
There have been attempts to connect optimal transport with spatiotemporal physics~\cite{hashimoto2026holographyoptimaltransportemergent}, mainly in the framework and gauge/gravity duality~\cite{hooft2009dimensionalreductionquantumgravity,Susskind1995,Maldacena1999,witten1998anti}. In this work, we apply our Kirkwood-Dirac phase space and spatiotemporal state framework to define quantum optimal transport in spacetime.

The Monge-Kantorovich optimal transport problem, originally proposed by Gaspard Monge in 1781~\cite{monge1781memoire}, concerns the optimal transportation of resources with respect to a prescribed cost function. 
A transport map $y=T(x)$ induces the push-forward $T_{\#}p$ of a probability measure $p(x)$ on $\cX$, with the constraint $T_{\#}p=q$ on $\cY$.
Given input and output probability measures $p$ and $q$, together with a cost function $c(x,T(x))$, Monge's problem is formulated as
\begin{equation}
    \operatorname{MP}
    =
    \inf\left\{
    \int_{\cX} c(x,T(x))p(x)\,dx
    \,\middle|\,
    T_{\#}p=q
    \right\}.
\end{equation}
A major advance was later made by Leonid Kantorovich~\cite{kantorovich2006translocation}, leading to the formulation now known under both of their names.
In Kantorovich's formulation, one is given input and output probability measures $p(x)$ and $q(y)$, along with a cost function $c(x,y)$ defined on the product polish space $\mathcal{X} \times \mathcal{Y}$. Imagine $\mathcal{X}$ as the producers and $\mathcal{Y}$ as the consumers; then $c(x,y)$ can be regarded as the cost of transporting goods from $x$ to $y$.
Transport plans are represented by joint probability distributions $\gamma(x,y)$ whose marginals coincide with $p(x)$ and $q(y)$, $\gamma(x,y)$ is also called coupling, the set of of coupling can be denoted as $\mathfrak{C}(p,q)$.   
The objective is to minimize the expected cost over all admissible couplings:
\begin{equation}
 \operatorname{KP} =  \inf_{\gamma\in \mathfrak{C}(p,q)} 
    \int_{\mathcal{X}\times \mathcal{Y}} c(x,y) \, \gamma(x,y) \, dx \, dy,
\end{equation}
For $k\in [1,\infty]$, Wasserstein metric is defined as~\cite{vaserstein1969markov} 
\begin{equation}
\begin{aligned}
       W_k(p,q)=&\inf_{\gamma\in \mathfrak{C}(p,q)} [\mathbb{E}_{\gamma(x,y)}d(x,y)^k]^{1/k} \\
       =&\inf_{\gamma\in \mathfrak{C}(p,q)} \left(\int_{\cX\times \cY} d(x,y)^k d\gamma(x,y) \right)^{1/k}
\end{aligned}
\end{equation}
The optimal transport problem has many applications in economics, fluid dynamics, meteorology and oceanography, surface reconstruction, etc., see \cite{villani2009optimal}.

Quantum generalizations of the Monge-Kantorovich problem have been studied from various perspectives since the 1990s.
Early work defined a distance between two quantum states as the Monge distance between their corresponding Husimi Q-functions \cite{zyczkowski1998monge,zyczkowski2001monge}. Replacing classical marginal distributions with density operators and classical couplings with joint density operators leads to a widely studied formulation of quantum optimal transport \cite{Golse2016On,Winter2016Tight,chakrabarti2019quantum,Zhou2022OPT,Caglioti2020Quantum,Friedland2020quantumOP,De2023Wasserstein,Toth2023quantumwasserstein,Cole2023On,Caglioti2023Towards,Caputo2025OPT}.
In this standard \emph{spatial} formulation, the prescribed marginals are local states $\rho_A$ and $\rho_B$, and a spatial quantum coupling is a spatial joint quantum state $\Ups_{AB}$ with reduced states $\rho_A$ and $\rho_B$. The objective is to minimize a prescribed quantum transport cost $\Tr [\sC \Ups_{AB}]$ over all such plans.
The quantum optimal transport also found many applications, like measuring distance between quantum states \cite{Zhou2022OPT,Friedland2020quantumOP,Cole2023On,DePalma2021Wasserstein,De2023Wasserstein} and quantum machine learning  \cite{DePalma2023OPT,chakrabarti2019quantum,Kiani2022OPT}.

Recently, some temporal generalizations have also been explored. In Ref.~\cite{DePalma2021QuantumMKchannal}, temporal quantum couplings are characterized by quantum channels, while in Ref.~\cite{hoogstederriera2026PDOMK}, pseudo-density operators are employed as temporal quantum transport plans. In this work, we propose a general framework for quantum optimal transport in spacetime using spatiotemporal quasiprobability distribution in spatiotemporal phase space and spatiotemporal states, which are systematically developed in \cite{Jia2025TemporalKirkwoodDirac,jia2026unifiedspatiotemporalquantumstates}, also see \cite{diaz2026unifyingspacetimeapproachesquantum,Guo2026Spacetime,milekhin2025observable,Lie2025stateovertime} for related discussions on spatiotemporal formalism of quantum mechanics.

Our main tool is the spatiotemporal Kirkwood-Dirac distribution \cite{Dirac1945on,Kirkwood1933quantum,Jia2025TemporalKirkwoodDirac,jia2026unifiedspatiotemporalquantumstates}, which can be regarded as a spatiotemporal state in the framework of generalized probabilistic theory.
In the spatiotemporal phase space, the local marginal states are characterized by quasiprobability distributions $Q_{z_i}(\alpha_i)$ over the phase space $\Omega_{z_i}$ for a given spacetime point $z_i=(x_i,t_i)$, the spatiotemporal coupling  is given by the joint phase-space distribution $Q_{\cA}(\boldsymbol{\alpha})$ over the joint phase space $\Omega_{\cA}=\Omega_{z_1}\times \cdots \times \Omega_{z_N}$ for the given spacetime set $\cA= \{ z_1,\cdots, z_N\}$. The cost function is a generally complex-valued function $C:\Omega_{\cA}\to \mathbb{C}$. The objective function is $\mid \sum_{\boldsymbol{\alpha}} C(\boldsymbol{\alpha}) Q(\boldsymbol{\alpha}) \mid$. The phase-space optimal transport is to find the infimum of  objective function. This is a very natural generalization of classical optimal transport, since one simply replaces the classical probability distribution with a quantum quasiprobability distribution; the quantumness is reflected in the deviation of the quasiprobability distribution from satisfying the Kolmogorov axioms.

Using frames $F_{\boldsymbol{\alpha}}$ and their dual frames $D_{\boldsymbol{\alpha}}$, we can transform the phase-space formulation into the spatiotemporal state formulation: the cost function becomes a cost operator $\sC=\sum_{\boldsymbol{\alpha}}C(\boldsymbol{\alpha})F_{\boldsymbol{\alpha}}$, and the Kirkwood-Dirac distribution yields the spatiotemporal state $\Ups=\sum_{\boldsymbol{\alpha}} Q(\boldsymbol{\alpha})D_{\boldsymbol{\alpha}}$. In general, both the cost operator and the spatiotemporal state can be non-Hermitian. The objective function becomes $\mid \Tr \sC \Ups \mid$. The standard spatial quantum optimal transport naturally becomes a special case of this problem if we choose $\cA$ to be an equal-time set $\cA=\{(x_1,t),\cdots,(x_N,t)\}$. Unlike the usual choice of a positive semidefinite cost operator, in this work we consider general non-Hermitian cost operators.

When the phase-space distribution is chosen as the Margenau-Hill distribution \cite{margenau1961correlation,Jia2025TemporalKirkwoodDirac,jia2026unifiedspatiotemporalquantumstates}, the corresponding spatiotemporal state becomes Hermitian; in this case, we obtain a Hermitian optimal transport problem.

The duality of optimal transport is established carefully; we discuss both Kantorovich duality and geometric duality. Defining the quantum Wasserstein metric is a crucial topic in quantum optimal transport. For the spatiotemporal case, directly defining a Wasserstein metric is challenging; instead, using the phase-space formalism, we can introduce the Kantorovich-Rubinstein metric, which can be regarded as a Wasserstein metric for $p=1$.

The rest of the paper is organized as follows. 
Section~\ref{sec:preliminary} reviews the fundamentals of the spatiotemporal phase space and Kirkwood-Dirac distributions over the phase space, and explains how it relates to the spatiotemporal state.
In Section~\ref{sec:QMKdef}, we define the quantum optimal transport using both spatiotemporal Kirkwood-Dirac distribution and spatiotemporal state. The phase-space definition can be transformed into the spatiotemporal-state definition via a frame and its dual frame; the reverse direction is given by the spatiotemporal Born rule.
We show that the spatiotemporal quantum optimal transport is independent of the choice of frames.
In Section~\ref{sec:spatial-KD-transport}, we apply our framework to the spatial setting, we especially carefully discuss the spatial optimal transport for Hermitian and non-Hermitian operators (both are beyond the standard positive semidefinite cost operator formalism). For Hermitian cost operator, we establish the quantum Kantorovich duality by showing that it is a duality of semidefinite programming (SDP). Using the phase-space distribution, this quantum Kantorovich duality has a very similar form as that for classical optimal transport.
For non-Hermitian operator, we show that the duality can be established via introducing an extra unit disk optimization in the complex plane. 
Section~\ref{sec:temporal-optimal-transport} applies the framework to the temporal setting. We carefully discuss the fixed-branch case and show that, in two-time case, it is equivalent to the quantum optimal transport for channels. For the multi-time case, it is a multi-convex problem.
Then we introduce the convex temporal coupling set for fixed-branch, and mixed-branch cases, and discuss how to establish the geometric duality for these problems.
We also discuss how to use the dual branch and Hermitianization to give a Margenau-Hill temporal quantum optimal transport.
In Section~\ref{sec:SToptoutline}, we briefly discuss how to extend our temporal optimal transport framework to the spatiotemporal setting.
Finally, in Section~\ref{sec:KRmetric}, we use the phase-space formalism to give a Kantorovich-Rubinstein metric by comparing it with the definition for classical optimal transport, we prove the metric satisfies the axioms of a distance function.
In Appendix, we collection some results from convex analysis and matrix analysis that is helpful for the discussion in the main text.

\section{Preliminaries on spatiotemporal quasiprobability distribution and its frame representation}
\label{sec:preliminary}

In this section, we will review the basics of spatiotemporal quasiprobability distributions and operator frames.
The standard Kirkwood-Dirac distribution is a complex-valued quasiprobabilistic representation of a spatial density operator with respect to a specified operator frame. Similarly, the spatiotemporal Kirkwood-Dirac distribution is a complex-valued quasiprobabilistic representation of a spatiotemporal density operator with respect to a specified operator frame. For a fixed frame, the spatiotemporal Kirkwood-Dirac distribution and the corresponding spatiotemporal density operator uniquely determine one another.

\subsection{Standard spatial Kirkwood-Dirac distribution}
To begin with, we recall several basic definitions. For a Hilbert space $\mathcal{H}$, we denote by $\mathbf{B}(\cH)$ the space of all bounded linear operators and by $\mathbf{Herm}(\cH)$ the space of all Hermitian operators. For an operator $A$, we denote its Schatten  $p$-norm by $\|A\|_p$. The Hilbert-Schmidt inner product is defined as $\langle A,B\rangle:=\Tr [A^{\dagger}B]$.
An operator frame~\cite{Duffin1952frame} is a family of operators $\{F_{\alpha}\}_{\alpha}$ for which there exist constants $0<a\le b<\infty$ such that, for every $A\in \mathbf{B}(\cH)$,
\begin{equation}
   a \|A\|_2^2\le \sum_{\alpha}\mid \Tr [F_{\alpha}^{\dagger}A] \mid^2 \le b\|A\|_2^2.
\end{equation}
A dual frame of $\{F_{\alpha}\}_{\alpha}$ is a family of operators $\{D_{\beta}\}_{\beta}$ satisfying the reconstruction formula
\begin{equation}
A=\sum_{\alpha}\Tr[F_{\alpha}^{\dagger}A]D_{\alpha},
\end{equation}
for every $A\in \mathbf{B}(\cH)$.
Operator frames play a central role in quasiprobabilistic descriptions of quantum theory; see Ref.~\cite{ferrie2011quasi} for a review.
Using an operator frame, a density operator $\rho$ can be represented by
\begin{equation}
    Q_{\rho}(\alpha)=\Tr [F_{\alpha}^{\dagger} \rho],
\end{equation}
and a measurement effect $E$ by
\begin{equation}
    \eta_E(\beta)=\Tr[D_{\beta}E].
\end{equation}
The Born rule then takes the form $\Tr [\rho E]= \sum_{\alpha }Q_{\rho}(\alpha) \eta_E(\alpha)$. In quantum theory, states and measurement effects cannot both admit classical probabilistic representations within a single frame representation; at least one must be represented quasiprobabilistically~\cite{Ferrie2010negativity}.

The Kirkwood-Dirac distribution admits a representation in terms of a non-Hermitian operator frame. Let $\{|a\rangle\}_{a=1}^d$ and $\{|b\rangle\}_{b=1}^d$ be orthonormal bases of $\cH$, with $P_a=|a\rangle\langle a|$ and $Q_b=|b\rangle\langle b|$, the standard  
Kirkwood-Dirac distribution for a density operator $\rho$ is given by \cite{Kirkwood1933quantum,Dirac1945on}
\begin{equation}
    Q(a,b)=\Tr [Q_bP_a\rho].
\end{equation}
Assume that $\langle b|a\rangle\neq0$ for every pair $(a,b)$. The operators
\begin{equation}
    F_{ab}=P_aQ_b
    =\langle a|b\rangle\,|a\rangle\langle b|
\end{equation}
form an orthogonal basis of $\mathbf{B}(\cH)$, since
\begin{equation}
    \Tr[F_{ab}^{\dagger}F_{a'b'}]
    =\delta_{aa'}\delta_{bb'}|\langle b|a\rangle|^2.
\end{equation}
To verify the frame condition explicitly, note that the operators $|a\rangle\langle b|$ form an orthonormal basis with respect to the Hilbert-Schmidt inner product. Therefore, for every $A\in\mathbf{B}(\cH)$,
\begin{align}
    \sum_{a,b}|\Tr[F_{ab}^{\dagger}A]|^2
    =\sum_{a,b}|\langle b|a\rangle|^2
    |\langle a|A|b\rangle|^2,\\
    \|A\|_2^2=\sum_{a,b}|\langle a|A|b\rangle|^2.
\end{align}
Defining $c_{\min}=\min_{a,b}|\langle b|a\rangle|^2>0$ and $c_{\max}=\max_{a,b}|\langle b|a\rangle|^2$, we obtain
\begin{equation}
    c_{\min}\|A\|_2^2
    \le \sum_{a,b}|\Tr[F_{ab}^{\dagger}A]|^2
    \le c_{\max}\|A\|_2^2.
\end{equation}
Thus $\{F_{ab}\}$ is an operator frame. It also satisfies $\sum_{a,b}F_{ab}=\mathds{1}$.
Its unique dual frame is
\begin{equation}
    G_{ab}
    =\frac{F_{ab}}{|\langle b|a\rangle|^2}
    =\frac{|a\rangle\langle b|}{\langle b|a\rangle},
\end{equation}
which satisfies $\Tr[F_{ab}^{\dagger}G_{a'b'}]=\delta_{aa'}\delta_{bb'}$.

The Kirkwood-Dirac distribution is the corresponding frame representation of a density operator:
\begin{equation}
    Q_{\rho}(a,b)
    =\Tr[F_{ab}^{\dagger}\rho]
    =\Tr[Q_bP_a\rho]
    =\langle b|a\rangle\langle a|\rho|b\rangle.
\end{equation}
The dual frame yields the reconstruction formula
\begin{equation}
    \rho=\sum_{a,b} Q_{\rho}(a,b)G_{ab}.
\end{equation}
A measurement effect $E$ is represented by
\begin{equation}
    \eta_E(a,b)
    =\Tr[G_{ab}E]
    =\frac{\langle b|E|a\rangle}{\langle b|a\rangle},
\end{equation}
so that the Born rule takes the form
\begin{equation}
    \Tr[\rho E]
    =\sum_{a,b}Q_{\rho}(a,b)\eta_E(a,b).
\end{equation}
Although the state and measurement representations can be complex, their pairing reproduces the physical measurement probability. Moreover, 
\begin{equation}
    \sum_bQ_{\rho}(a,b)=\Tr[P_a\rho],
    \sum_aQ_{\rho}(a,b)=\Tr[Q_b\rho],
\end{equation}
the Kirkwood-Dirac distribution has the measurement probabilities in the two bases as its marginals.

\subsection{Spatiotemporal Kirkwood-Dirac distribution}

A spatiotemporal Kirkwood-Dirac distribution is a complex-valued quasiprobability representation of a spatiotemporal density operator with respect to a chosen spatiotemporal operator frame~\cite{Jia2025TemporalKirkwoodDirac,jia2026unifiedspatiotemporalquantumstates}. We focus primarily on the temporal setting; the extension to the spatiotemporal setting is straightforward.

Following Refs.~\cite{Jia2025TemporalKirkwoodDirac,jia2026unifiedspatiotemporalquantumstates}, we define the temporal Kirkwood-Dirac distribution for memoryless dynamics as\footnote{For the non-Markovian case, we need to introduce an ancillary system and trace it out at the final step, see \cite{jia2026unifiedspatiotemporalquantumstates}.}
\begin{equation}
\begin{aligned}
  & Q(\alpha_N, \dots, \alpha_1) \\
=&
\operatorname{Tr} \Big[ \mathcal{P}^{(t_N)}_{\alpha_N} \circ \mathcal{E}_{t_n \leftarrow t_{n-1}} \circ \cdots \circ \mathcal{P}^{(t_1)}_{\beta_1} \circ \mathcal{E}_{t_1 \leftarrow t_0} \circ \mathcal{P}^{t_0}_{\beta_0} (\rho_{t_0}) \Big], 
\end{aligned}
\label{eq:temporal_quasiprobability}
\end{equation}
where $\cE_{t_{k+1} \leftarrow t_k}$ is the completely positive trace-preserving (CPTP) evolution map from $t_k$ to $t_{k+1}$, $\alpha_k$ labels a phase-space point at time $t_k$, and $\mathcal{P}^{(t_k)}_{\alpha_k}$ is the corresponding phase-space operation, which need not be completely positive. Typical choices~\cite{Jia2025TemporalKirkwoodDirac,jia2026unifiedspatiotemporalquantumstates} include right multiplication, $\mathcal{P}^{(t_k)}_{\alpha_k}(\bullet) = \bullet\, F^{(t_k)}_{\alpha_k}$; left multiplication, $\mathcal{P}^{(t_k)}_{\alpha_k}(\bullet) = F^{(t_k)}_{\alpha_k}\, \bullet$; and the doubled construction, $\mathcal{P}^{t_k}_{\alpha_k,\beta_k}(\bullet) = F^{(t_k)}_{\alpha_k}\, \bullet\, F^{(t_k)}_{\beta_k}$, where $\{F^{(t_k)}_{\alpha_k}\}$ and $\{F^{(t_k)}_{\beta_k}\}$ are operator frames. 
Following the convention introduced in \cite{jia2026unifiedspatiotemporalquantumstates}, we refer to the choice made at each time slot as a branch and denote $\mathtt{B}=(t_n^{s_n},\cdots, t_0^{s_0})$ with $s_k=L,R$.
See ~\cite{Jia2025TemporalKirkwoodDirac,jia2026unifiedspatiotemporalquantumstates} for comprehensive discussion.

\begin{figure}
    \centering
    \includegraphics[width=1\linewidth]{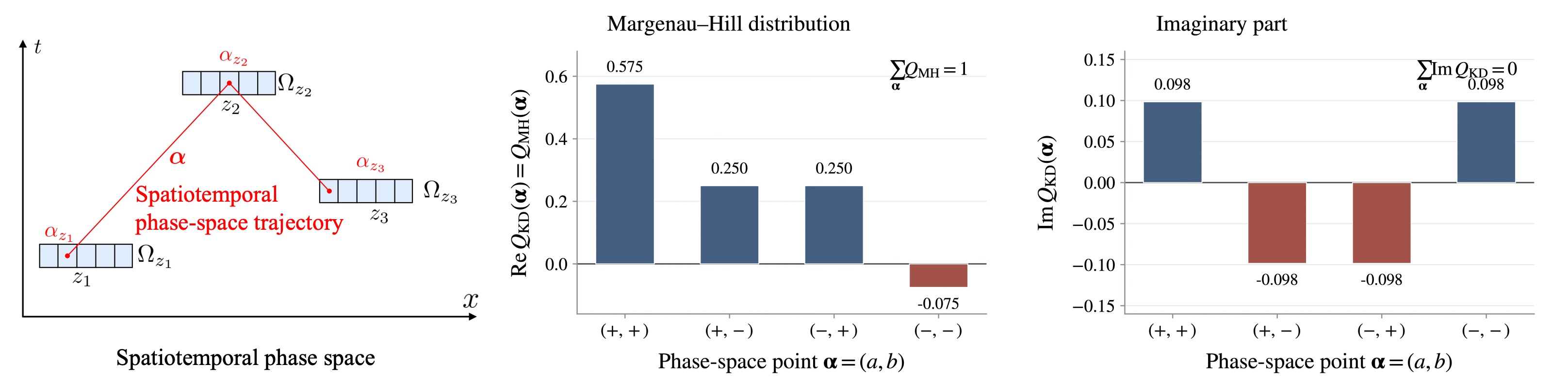}
    \caption{Illustration of spatiotemporal phase space and spatiotemporal Kirkwood-Dirac distribution. The real part of spatiotemporal Kirkwood-Dirac distribution is a real-valued quasiprobability distribution called Margenau-Hill quasiprobability distribution.}
    \label{fig:STphasespace}
\end{figure}

For spacetime lattice $\cA=\{(t_1,x_1),\cdots, (t_N,x_N)\}$ with the associated spatiotemporal Hilbert space $\mathcal{H}_{\cA}=\bigotimes_{k}\cH_{(t_k,x_k)}$, a spatiotemporal state is an operator of the form
\begin{equation}
    \Upsilon_{\cA} = \frac{1}{d^N} \sum_{\mu_1,\cdots,\mu_N} T^{\mu_1,\cdots,\mu_N} \sigma_{\mu_1}\otimes \cdots \otimes \sigma_{\mu_N}.
\end{equation}
Here, $\sigma_k$ are generalized Gell-Mann matrices, and $T^{\mu_1,\cdots,\mu_N}$ is the spatiotemporal joint expectation value of $\sigma_{\mu_1}(t_1,x_1)$, $\cdots$, $\sigma_{\mu_N}(t_N,x_N)$.
Notice that $\Upsilon_{\cA}$ is generally non-Hermitian and non-normal, but all reduced state $\Upsilon_{(t_k,x_k)}$ is a density operator.
The spatiotemporal Kirkwood-Dirac distribution provides a quasiprobability representation of $\Upsilon$ with respect to a chosen operator frame~\cite{Jia2025TemporalKirkwoodDirac,jia2026unifiedspatiotemporalquantumstates},
\begin{equation}
    Q(\alpha_1,\cdots,\alpha_N) =\Tr [(F_{\alpha_1}^{(1)}\otimes \cdots \otimes F_{\alpha_N}^{(N)})\Upsilon]
\end{equation}
with $F_{\alpha_k}$'s frames.
Thus, the spatiotemporal Kirkwood-Dirac distribution can be viewed as a quasiprobabilistic spatiotemporal state within the framework of generalized probabilistic theory.
Write $z_k=(t_k,x_k)$ and let $\mathcal{H}_k$ be the Hilbert space
associated with $z_k$. We use the notation
\begin{equation}
\boldsymbol{\alpha}=(\alpha_1,\ldots,\alpha_N),
\quad
F_{\boldsymbol{\alpha}}
=
F^{(1)}_{\alpha_1}\otimes\cdots\otimes F^{(N)}_{\alpha_N}.
\end{equation}
The superscript on $F^{(k)}_{\alpha_k}$ indicates the spacetime site on
which the operator acts. We call
$\boldsymbol{\alpha}$  a spatiotemporal phase-space point, the set of them is the spatiotemporal phase space. See Figure~\ref{fig:STphasespace} for an illustration.
For the standard Kirkwood-Dirac distribution, each local phase-space point is a pair $\alpha_i=(a_i,b_i)$.
Characterizing the full set of Kirkwood-Dirac distributions arising from density operators is generally challenging; partial results have been obtained, see~\cite{DeBievre2021KD,debivre2026kirkwooddirac} and references therein. For spatiotemporal case. the problem  largely remain open.

Recall that a positive operator-valued measure (POVM) is a family of positive semidefinite  operators $\{F_{\alpha}\}_{\alpha}\subset \mathbf{Herm}_{\ge 0}(\cH)$ satisfying $\sum_{\alpha}F_{\alpha}=\mathds{1}$. A POVM is informationally complete (IC) if its elements span the real vector space of Hermitian operators, $\operatorname{span}_{\mathbb{R}} \{F_{\alpha}\} =\Herm (\cH)$. Since every linear operator is a complex linear combination of Hermitian operators, an IC-POVM also spans the complex vector space of all linear operators, $\operatorname{span}_{\mathbb{C}} \{F_{\alpha}\} =\mathbf{B} (\cH)$. If $\dim \cH=d$, an IC-POVM with $d^2$ elements is called minimal.
In finite dimensions, a POVM forms an operator frame if and only if it is IC. Throughout this work, we restrict attention to finite-dimensional Hilbert spaces and take our frames to be minimal IC-POVMs.
In this case the dual frame $G_{\beta}$ is given by
\begin{equation}\label{eq:dualframe}
    D_{\beta}=\sum_{\alpha} G^{-1}_{\alpha\beta}F_{\alpha},
\end{equation}
where $G_{\alpha\beta}=\Tr [F_{\alpha}F_{\beta}]$ is Gram matrix and we have $\Tr [F_{\alpha}D_{\beta}]=\delta_{\alpha\beta}$. For over-complete POVM frame, we need replace the inverse in Eq.~\eqref{eq:dualframe} with Moore-Penrose pseudoinverse.

For two minimal IC-POVM frames \(\{F_\alpha\}\) and \(\{F'_\alpha\}\) with dual frames \(\{D_\alpha\}\) and \(\{D'_\alpha\}\), suppose that they are related by an invertible transformation \(M\):
\begin{equation}
    F'_\alpha = \sum_\beta M_{\alpha\beta} F_\beta .
\end{equation}
Then the dual frames transform as
\begin{equation}
    D'_\alpha = \sum_\beta (M^{-1})_{\beta\alpha} D_\beta .
\end{equation}

At each spacetime point $z_k$, we can choose an IC-POVM as the operator frame. The resulting single-time distribution $Q(\alpha_k)=\Tr [F_{\alpha_k}\rho_{t_k}]$ is then a probability distribution. As we will see, this choice is particularly convenient for formulating the spatiotemporal quantum optimal transport problem in phase space, since it is a natural generalization of that for classical phase space.

\section{Defining quantum optimal transport in spacetime}
\label{sec:QMKdef}

In this section, we define the quantum Monge-Kantorovich optimal transport problem in spacetime in the general setting, which corresponds to the multi-marginal optimal transport problem in classical theory. We refer the reader to \cite{villani2009optimal} for basic definitions and results of optimal transport.

Consider a spacetime set $\cA=\{(t_1,x_1),\cdots, (t_N,x_N)\}$  and the corresponding local minimal IC-POVM frames\footnote{This choice is made for convenience; the quantum Monge-Kantorovich optimal transport problem in spatiotemporal phase space can also be defined using general operator frames.} $\{F_{\alpha_1}\}$, $\cdots$, $\{F_{\alpha_N}\}$.  Set $z_k=(t_k,x_k)$, suppose that we are given local density operators $\rho_{z_k}$ and the associated local quantum distributions $Q_{z_k}(\alpha_k)=\Tr [F_{\alpha_k}\rho_{z_k}]$. We denote the collection of these local marginal distributions by $\mathcal{Q}_{\cA}$. A joint spatiotemporal Kirkwood-Dirac distribution $Q_{\cA}(\alpha_1,\cdots,\alpha_N)$ with marginals $Q_{z_k}(\alpha_k)$ can be regarded as a \emph{spatiotemporal coupling}. We denote the set of all such Kirkwood-Dirac spatiotemporal couplings by $\mathbb{KD}(\mathcal{Q}_{\cA})$. 
The Margenau-Hill spatiotemporal couplings is the real part of Kirkwood-Dirac spatiotemporal couplings, we similarly denote their set as $\mathbb{MH}(\mathcal{Q}_{\cA})$.

For a given cost function $C(\alpha_1,\cdots,\alpha_N)$, the quantum Monge-Kantorovich optimal transport problem in spatiotemporal phase space is defined as follows
\begin{equation}\label{eq:KDoptimal}
   \operatorname{QMK}_{\rm KD}= \inf_{Q_{\cA}\in\mathbb{KD}(\mathcal{Q}_{\cA})}
    \left|\sum_{\alpha_1,\cdots,\alpha_N}
    C(\alpha_1,\cdots,\alpha_N)\,
    Q_{\cA}(\alpha_1,\cdots,\alpha_N)\right|.
\end{equation}
Note that ``all such couplings'' requires further clarification; specifically, one must specify the physical Kirkwood-Dirac distributions in a chosen frame. As we will see, they are equivalent to the quantum optimal transport based on spatiotemporal state, since Kirkwood-Dirac distribution is obtained from spatiotemporal state via spatiotemporal Born rule \cite{Jia2025TemporalKirkwoodDirac,jia2026unifiedspatiotemporalquantumstates}. We will give more details later about the spatiotemporal distributions and spatiotemporal  states where the optimization is taken over.

For a given set of density operators $\{\rho_{z_k}\}_{z_k\in \cA}$, the set of multi-marginal quantum spatiotemporal state couplings is
\begin{equation}
\mathfrak{C}(\rho_{z_1},\cdots,\rho_{z_N})
=
\left\{
\Upsilon_{\cA}\in \mathbf{State}_{\rm ST}(\cH_{\cA}): \operatorname{Tr}_{\widehat{z_k}}\Upsilon_{\cA}
=
\rho_{z_k} \forall k
\right\},
\label{eq:spacetime-quantum-couplings}
\end{equation}
where $\operatorname{Tr}_{\widehat{z_k}}$ denotes the partial trace over
all tensor factors except the $k$-th one and $\mathbf{State}_{\rm ST}(\cH_{\cA})$ denote the set of all spatiotemporal density operators.
This set is nonempty since
\begin{equation}
\rho_{z_1}\otimes\cdots\otimes\rho_{z_N}
\in
\fC(\rho_{z_1},\ldots,\rho_{z_N}),
\end{equation}
recall the we can use replacement channel as evolution to obtain the spatiotemporal product state.

Every spatiotemporal state $\Upsilon_{\cA}$ defines a joint spatiotemporal phase-space distribution via the Born rule \cite{Jia2025TemporalKirkwoodDirac,jia2026unifiedspatiotemporalquantumstates}
\begin{equation}\label{eq:KDfromSTstate}
Q_{\Upsilon_{\cA}}(\boldsymbol{\alpha})
=
\operatorname{Tr}
\left[
F_{\boldsymbol{\alpha}}\Upsilon_{\cA}
\right],
\end{equation}
which is the quasiprobabilistic representation of the spatiotemporal state in the chosen frame. Its $k$-th marginal is given by
\begin{align}
\sum_{\alpha_1,\ldots,\widehat{\alpha_k},\ldots,\alpha_N}
Q_{\Upsilon_{\cA}}(\boldsymbol{\alpha})
&=
\operatorname{Tr}
\left[
F^{(k)}_{\alpha_k}
\operatorname{Tr}_{\widehat{k}}\Upsilon_{\cA}
\right]
\nonumber\\
&=
\operatorname{Tr}
\left[
F^{(k)}_{\alpha_k}\rho_{z_k}
\right]
=
Q_{z_k}(\alpha_k).
\label{eq:joint-spacetime-marginals}
\end{align}
Here the hat over $\alpha_k$ indicates that this index is not summed over. For each local IC-POVM, we choose a dual operator frame
$\{D^{(k)}_{\alpha_k}\}$ satisfying
\begin{equation}
X
=
\sum_{\alpha_k}
\operatorname{Tr}
\left[
F^{(k)}_{\alpha_k}X
\right]
D^{(k)}_{\alpha_k}
\end{equation}
for every operator $X$ on $\mathcal{H}_{z_k}$. Given a spatiotemporal Kirkwood-Dirac distribution
$Q_{\cA}(\boldsymbol{\alpha})$, we define its reconstructed spatiotemporal state by
\begin{equation}\label{eq:STreconstruct}
\Upsilon_{Q_{\cA}}
=
\sum_{\alpha_1,\ldots,\alpha_N}
Q_{\cA}(\boldsymbol{\alpha})\,
D^{(1)}_{\alpha_1}
\otimes\cdots\otimes
D^{(N)}_{\alpha_N}.
\end{equation}
The set of physical Kirkwood-Dirac spatiotemporal couplings is then
\begin{align}
\mathbb{KD}(\mathcal{Q}_{\cA})
=
\Bigl\{
Q:\;&
\Upsilon\in \mathbf{State}_{\rm ST}(\cH_{\cA}),
\quad
Q(\boldsymbol{\alpha})
=
\operatorname{Tr}
\left[
F_{\boldsymbol{\alpha}}\Upsilon_Q
\right],
\nonumber\\
&
\sum_{\alpha_1,\ldots,\widehat{\alpha_k},\ldots,\alpha_N}
Q(\boldsymbol{\alpha})
=
Q_{z_k}(\alpha_k)
\text{ for every }k
\Bigr\}.
\label{eq:quantum-compatible-spacetime-plans}
\end{align}
When each local IC-POVM is minimal, fixing the local $Q_{z_k}(\alpha_k)$ distribution is equivalent to fixing the local density operators $\rho_{z_k}$.

From the perspective of generalized probabilistic theory, we can also define the optimal transport beyond quantum theory.
Let $\mathbb{QD}_r^{\Cbb}(\mathcal{Q}_{\cA})$ denote the set of all complex-valued quasiprobability distribution with fixed marginals that take values in disk $\mathbb{D}=\{w
\in \Cbb: |w|\leq r\}$, for sufficiently large $r$,
\begin{equation}
\mathbb{KD}(\mathcal{Q}_{\cA})
\subset \mathbb{QD}_r^{\Cbb}(\mathcal{Q}_{\cA})
\end{equation}
and this inclusion is generally strict. If we allow the coupling to be any elements in $\mathbb{QD}_r^{\Cbb}(\mathcal{Q}_{\cA})$, we will obtain a optimal transport problem
\begin{equation}
 \operatorname{QMK}_{\rm QD_r^{\Cbb}} =   \inf_{Q_{\cA}\in\mathbb{QD}(\mathcal{Q}_{\cA})}
    \left|
    \sum_{\alpha_1,\cdots,\alpha_N}
    C(\alpha_1,\cdots,\alpha_N)\,
    Q_{\cA}(\alpha_1,\cdots,\alpha_N)
    \right|.
\end{equation}
We have $ \operatorname{QMK}_{\rm QD_r^{\Cbb}} \le \operatorname{QMK}_{\rm KD}$ for sufficiently large $r$. 
When restricting to the real quasiprobability distributions and real cost functions, the problem reduce to the classical optimal transport problem for signed measures.
The real quasiprobability distributions coupling also has physical subset of Margenau-Hill distributions $\mathbb{MH}(\mathcal{Q}_{\cA})$, there will also be a corresponding spatiotemporal optimal transport problem $\operatorname{QMK}_{\rm MH}$. Similarly, $\operatorname{QMK}_{\rm MH} \ge \operatorname{QMK}_{\rm QD^{\Rbb}_r}$ for sufficiently large $r$.

The phase-space cost function determines the cost operator and the cost operator has a unique frame expansion in minimal IC-POVM frame
\begin{equation}
\mathsf{C}
=
\sum_{\boldsymbol{\alpha}}
C_{\mathsf{C}}(\boldsymbol{\alpha})
\bigotimes_{k=1}^N F^{(k)}_{\alpha_k},
\quad
C_{\mathsf{C}}(\boldsymbol{\alpha})
=
\Tr\left[
\left(\bigotimes_{k=1}^N D^{(k)}_{\alpha_k}\right)
\mathsf{C}
\right].
\label{eq:spacetime-cost-operator}
\end{equation}
Using the spatiotemporal state, we can also define the spatiotemporal state optimal transport as
\begin{equation}
    \inf_{\Upsilon\in \mathfrak{C}(\rho_{z_1},\ldots,\rho_{z_N})} \left| \Tr[\mathsf{C} \Upsilon] \right|,
\end{equation}
which is equivalent to the phase-space formulation in Eq.~\eqref{eq:KDoptimal}.

\begin{theorem}[Equivalence of spatiotemporal phase-space optimal transport and spatiotemporal state optimal transport]\label{thm:KD-OPT}
  For a spacetime set $\cA$ and minimal IC-POVM frames, the map given by Eq.~\eqref{eq:KDfromSTstate}
\begin{equation}
\Upsilon_{\cA}
\longmapsto
Q_{\Upsilon_{\cA}}
\end{equation}
is a bijection from
$\mathfrak{C}(\rho_{z_1},\ldots,\rho_{z_N})$ onto
$\mathbb{KD}(\mathcal{Q}_{\cA})$, with inverse
given in Eq.~\eqref{eq:STreconstruct}.
In the chosen frame, we have
\begin{equation}
    \Tr [\sC\Upsilon]=\sum_{\alpha_1,\cdots,\alpha_N} C(\alpha)Q(\mathbf{\alpha})
\end{equation}
where $Q(\mathbf{\alpha})$ and $C(\alpha)$ are coefficients in the frame and dual frame. Thus the phase-space optimal transport is equivalent to the spatiotemporal state optimal transport.
\end{theorem}

\begin{proof}
    Informational completeness of the local frames implies informational
completeness of their tensor product. Consequently,
\begin{equation}
X
=
\sum_{\boldsymbol{\alpha}}
\operatorname{Tr}
\left[
F_{\boldsymbol{\alpha}}X
\right]
D^{(1)}_{\alpha_1}
\otimes\cdots\otimes
D^{(N)}_{\alpha_N}
\label{eq:multisite-frame-reconstruction}
\end{equation}
for every operator $X$ on
$\mathcal{H}_{z_1}\otimes\cdots\otimes\mathcal{H}_{z_N}$. This proves
injectivity and shows that
$\Upsilon_{Q_{\Upsilon_{\cA}}}=\Upsilon_{\cA}$.

If $\Upsilon_{\cA}$ has the required reduced states, then
\eqref{eq:joint-spacetime-marginals} gives the prescribed marginals.
Conversely, let $Q$ satisfy
\eqref{eq:quantum-compatible-spacetime-plans}. For every $k$,
\begin{equation}
\operatorname{Tr}
\left[
F^{(k)}_{\alpha_k}
\operatorname{Tr}_{\widehat{k}}\Upsilon_Q
\right]
=
\operatorname{Tr}
\left[
F^{(k)}_{\alpha_k}\rho_{z_k}
\right].
\end{equation}
Informational completeness gives
$\operatorname{Tr}_{\widehat{k}}\Upsilon_Q=\rho_{z_k}$, it follows that
$\Upsilon_Q\in\mathfrak{C}(\rho_{z_1},\ldots,\rho_{z_N})$.
\end{proof}

Note that in Eq.~\eqref{eq:KDoptimal} we choose the absolute value as the objective function; however, there are many other choices we can consider:
\begin{itemize}
    \item Absolute value \(r_C(Q) := \bigl|\sum_{\boldsymbol{\alpha}} C(\boldsymbol{\alpha}) Q(\boldsymbol{\alpha})\bigr| = \bigl|\operatorname{Tr}[\mathsf{C}\,\Upsilon]\bigr| =: r_{\mathsf{C}}(\Upsilon)\).
    \item Phase value \(\theta_C(Q) := -i \operatorname{Log}\!\left( \frac{\sum_{\boldsymbol{\alpha}} C(\boldsymbol{\alpha}) Q(\boldsymbol{\alpha})}{\bigl|\sum_{\boldsymbol{\alpha}} C(\boldsymbol{\alpha}) Q(\boldsymbol{\alpha})\bigr|} \right) = -i \operatorname{Log}\!\left( \frac{\operatorname{Tr}[\mathsf{C}\,\Upsilon]}{\bigl|\operatorname{Tr}[\mathsf{C}\,\Upsilon]\bigr|} \right) =: \theta_{\mathsf{C}}(\Upsilon)\).
    \item Real part \(u_C(Q) = \operatorname{Re} \sum_{\boldsymbol{\alpha}} C(\boldsymbol{\alpha}) Q(\boldsymbol{\alpha}) = \operatorname{Re} \operatorname{Tr}[\mathsf{C}\,\Upsilon] =: u_{\mathsf{C}}(\Upsilon)\).
    \item Imaginary part \(v_C(Q) = \operatorname{Im} \sum_{\boldsymbol{\alpha}} C(\boldsymbol{\alpha}) Q(\boldsymbol{\alpha}) = \operatorname{Im} \operatorname{Tr}[\mathsf{C}\,\Upsilon] =: v_{\mathsf{C}}(\Upsilon)\).
\end{itemize}
We will mainly focus on the absolute value and real part cases, since they are directly related the those considered in spatial quantum optimal transport.

For a general operator \(X\), define its Hermitian components by
\begin{equation}
    X_{\mathrm{R}} = \frac{X + X^{\dagger}}{2},
    \quad
    X_{\mathrm{I}} = \frac{X - X^{\dagger}}{2i}.
    \label{eq:real-imaginary-cost}
\end{equation}
When \(X \in \Herm(\mathcal{H})\), we have \(X_{\mathrm{R}} = X\) and \(X_{\mathrm{I}} = 0\).
For every \(\Upsilon \in \mathbf{State}_{\mathrm{ST}}\) and cost operator \(\mathsf{C}\), we have
\begin{equation}
    \operatorname{Tr}[\mathsf{C}\,\Upsilon]
    = \operatorname{Tr}[\mathsf{C}_{\mathrm{R}} \Upsilon_{\mathrm{R}} - \mathsf{C}_{\mathrm{I}} \Upsilon_{\mathrm{I}}]
    + i \operatorname{Tr}[\mathsf{C}_{\mathrm{I}} \Upsilon_{\mathrm{R}} + \mathsf{C}_{\mathrm{R}} \Upsilon_{\mathrm{I}}].
\end{equation}
When \(\mathsf{C} \in \mathbf{Herm}(\mathcal{H})\), \(\operatorname{Re} \operatorname{Tr}[\mathsf{C}\,\Upsilon] = \operatorname{Tr}[\mathsf{C}\,\Upsilon_{\mathrm{R}}]\). Thus, for a Hermitian cost operator and a real-part objective function, the Kirkwood-Dirac quantum optimal transport reduces to the Margenau-Hill quantum optimal transport. When \(\Upsilon \in \mathbf{Herm}(\mathcal{H})\), namely in the Margenau-Hill phase space, we have \(\operatorname{Re} \operatorname{Tr}[\mathsf{C}\,\Upsilon] = \operatorname{Tr}[\mathsf{C}_{\mathrm{R}}\,\Upsilon]\). When both \(\mathsf{C}\) and \(\Upsilon\) are Hermitian, we have \(\operatorname{Re} \operatorname{Tr}[\mathsf{C}\,\Upsilon] = \operatorname{Tr}[\mathsf{C}\,\Upsilon]\). This means that in the above three cases we always obtain the Margenau-Hill quantum optimal transport. The real-part objective function also gives the correct classical limit, whereas the absolute value function introduces an extra absolute value sign.

\subsection{Optimal transport in projector-product Kirkwood-Dirac phase space}
\label{subsec:KDopt}

The previous discussion mainly use the IC-POVM frames, we can also define the optimal transport using 
projector-product frame.
For $\cH_{z_k}$, let
$\{|a_k\rangle\}$ and $\{|a'_k\rangle\}$ be two orthonormal bases with nonzero mutual overlaps, and write
\begin{equation}
P_{a_k}=|a_k\rangle\langle a_k|,
\quad
\widetilde{P}_{a'_k}=|a'_k\rangle\langle a'_k|, \quad K^A_{a_ka'_k}=P_{a_k}\widetilde{P}_{a'_k}.
\end{equation}
The local Kirkwood-Dirac distributions are
\begin{equation}
Q_{\rho_{z_k}}(a_k,a'_k)
=
\Tr\left[
(K_{a_ka'_k})^\dagger\rho_{z_k}
\right],
\label{eq:local-standard-KD}
\end{equation}
which, unlike those for IC-POVM frames, are complex-valued quasiprobability distributions.
The projector-product Kirkwood-Dirac couplings are $\mathbb{KD}^{\mathrm{proj}}
(\mathcal{Q}_{\cA})$.
For cost function
$C(a_1,a'_1;\cdots;a_N,a'_N)$, define
\begin{equation}
\mathsf{C}
=
\sum_{a_1,a'_1;\cdots;a_N,a'_N}
C(a_1,a'_1;\cdots;a_N,a'_N)
(K^A_{a_1,a'_1})^\dagger
\otimes \cdots \otimes
(K^B_{a_N,a'_N})^\dagger.
\label{eq:projector-KD-cost-operator}
\end{equation}
Then
\begin{align}
&
\inf_{Q\in
\mathbb{KD}^{\mathrm{proj}}
(\mathcal{Q}_{\cA})}
\left|
\sum_{a_1,a'_1;\cdots;a_N,a'_N}
C(a_1,a'_1;\cdots;a_N,a'_N)Q(a_1,a'_1;\cdots;a_N,a'_N)
\right|
\nonumber\\
&\quad=
\inf_{\Ups\in\mathfrak{C}(\rho_{z_1},\cdots, \rho_{z_N})}
\left|\Tr(\mathsf{C}\Ups)\right|.
\label{eq:projector-KD-MK-equivalence}
\end{align}

\section{Spatial optimal transport in spatial phase space}
\label{sec:spatial-KD-transport}

We now specialize the preceding spatiotemporal construction to a purely
spatial setting $\cA=(x_1,\cdots,x_N)$. This specialization identifies the
phase-space formulation with the density-operator formulation of
 quantum optimal transport and clarifies which features
are specifically spatiotemporal. The main difference is that a spatial transport plan is a positive
multipartite density operator, whereas a general spatiotemporal state
need not be positive.

Let $\mathcal{H}_{x_1},\ldots,\mathcal{H}_{x_N}$ be finite-dimensional Hilbert
spaces, let
$\mathcal{H}_{\cA}=\bigotimes_{k=1}^N\mathcal{H}_{x_k}$, and fix density operators $\rho_k$ on each $\mathcal{H}_{x_k}$. Write
$\widehat{x_k}=\{x_1,\ldots,x_N\}\setminus\{x_k\}$, so that
$\Tr_{\widehat{x}_k}$ traces out every subsystem except $k$.
The set of spatial quantum couplings with these marginals is
\begin{equation}
\mathfrak{C}(\rho_1,\ldots,\rho_N)
=
\left\{
\Ups_{\cA}\geq0:
\Tr\Ups_{\cA}=1,
\Tr_{\widehat{x_k}}\Ups_{\cA}=\rho_k
\text{ for every }k
\right\}.
\label{eq:spatial-quantum-couplings}
\end{equation}
This set is nonempty, compact, and convex; in particular,
$\bigotimes_{k=1}^N\rho_k\in
\mathfrak{C}(\rho_1,\ldots,\rho_N)$.

Fix minimal IC-POVM frame and phase space $\Omega_{\cA}=\prod_{k=1}^N \Omega_{x_k}$, every spatial coupling $\Ups_{\cA}$ has the joint
phase-space representation
\begin{equation}
Q_\Ups(\boldsymbol{\alpha})
=
\Tr\left[
F_{\boldsymbol{\alpha}}\Ups_{\cA}
\right].
\label{eq:joint-spatial-IC-distribution}
\end{equation}
It satisfies
\begin{equation}
Q_\Ups(\boldsymbol{\alpha})\geq0,
\quad
\sum_{\boldsymbol{\alpha}}Q_\Ups(\boldsymbol{\alpha})=1,
\label{eq:joint-IC-positivity}
\end{equation}
and its single-subsystem marginals are
\begin{equation}
\sum_{\boldsymbol{\alpha}_{\widehat{k}}}
Q_\Ups(\boldsymbol{\alpha})
=
Q_{x_k}(\alpha_k)=\Tr [F_{\alpha_k}\rho_{k}],
\quad k=1,\ldots,N,
\label{eq:joint-IC-marginals}
\end{equation}
where the sum runs over all outcome labels except $\alpha_k$.

An arbitrary classical coupling of $Q_{x_1},\ldots,Q_{x_N}$ need not arise
from a spatial quantum coupling. 
Accordingly, we define the physical spatial phase-space transport
coupling as
\begin{equation}
\begin{split}
\mathbb{KD}_{\mathrm{sp}}(p_1,\ldots,p_N)
=
\Bigl\{
Q:\Omega_{\cA}\longrightarrow\mathbb{R}:\;&
\sum_{\boldsymbol{\alpha}_{\widehat{k}}}
Q(\boldsymbol{\alpha})=p_k(\alpha_k)
\forall k,
\Ups_Q\in \mathbf{State}_{\rm S}
\Bigr\}.
\end{split}
\label{eq:physical-spatial-KD-plans}
\end{equation}
We can equivalently replace condition $\Ups_Q\in \mathbf{State}_{\rm S}$ with $\Ups_Q\ge 0$. The marginal conditions imply $\Tr\Ups_Q=1$.
Thus, within the set of classical couplings, quantum realizability
is characterized by the semidefinite constraint $\Ups_Q\geq0$.

In particular, for any real-valued
cost function $C:\Omega_{\cA}\to\mathbb{R}$, define cost operator as that in Eq.~\eqref{eq:spacetime-cost-operator}.
The Theorem~\ref{thm:KD-OPT} shows that spatial phase-space optimal transport is equivalent to joint density operator optimal transport
\begin{equation}
\inf_{Q\in\mathbb{KD}_{\mathrm{S}}(Q_{x_1},\ldots,Q_{x_N})}
\sum_{\boldsymbol{\alpha}}
C(\boldsymbol{\alpha})Q(\boldsymbol{\alpha})
=
\inf_{\Ups_{\cA}\in
\mathfrak{C}(\rho_{1},\ldots,\rho_{N})}
\Tr(\sC\Ups_{\cA}).
\label{eq:spatial-multimarginal-transport-equivalence}
\end{equation}
For a nonnegative cost function, $\sC\geq0$, it reduces to the quantum optimal transport discussed in \cite{Golse2016On,Winter2016Tight,chakrabarti2019quantum,Zhou2022OPT,Caglioti2020Quantum,Friedland2020quantumOP,De2023Wasserstein,Toth2023quantumwasserstein,Cole2023On,Caglioti2023Towards,Caputo2025OPT}.

Conversely, every cost operator $\mathsf{C}$ on
$\mathcal{H}_{\cA}$ has the unique frame expansion given in Eq.~\eqref{eq:spacetime-cost-operator}.
If $\mathsf{C}$ is Hermitian, this kernel is real. However,
$\mathsf{C}\geq0$ does not generally imply
$C(\boldsymbol{\alpha})=\Tr D_{\boldsymbol{\alpha}} \sC\geq0$, because the dual-frame
elements need not be positive. Nevertheless, positivity of
$\mathsf{C}$ ensures
$\Tr(\mathsf{C}\Ups_{\cA})\geq0$ for every physical
coupling.

\begin{proposition}[Frame independence]
\label{prop:frame-covariance}
Fix a set of local density operators $\rho_1,\ldots,\rho_N$, consider two collections of local minimal IC-POVM frames, the corresponding joint
phase-space distributions are related by an invertible linear
transformation, while the cost function transform
contragrediently. The optimal Monge-Katorovich value is unchanged.
\end{proposition}

\begin{proof}
Write the change of local frames as
\begin{equation}
\widetilde F^{(k)}_{\mu_k}
=
\sum_{\alpha_k}
M^{(k)}_{\mu_k\alpha_k}F^{(k)}_{\alpha_k},
\end{equation}
where each $M^{(k)}$ is a real invertible matrix, and set
$M=\bigotimes_{k=1}^N M^{(k)}$.
Viewing the distributions and cost functions as column vectors, the
joint distributions satisfy
\begin{equation}
\widetilde Q_\Ups=M Q_\Ups.
\end{equation}
To represent the same cost operator, the kernels must satisfy
\begin{equation}
\widetilde C_{\mathsf{C}}
=
(M^{-1})^{T}C_{\mathsf{C}}.
\end{equation}
Consequently,
\begin{equation}
\sum_{\boldsymbol{\mu}}
\widetilde C_{\mathsf{C}}(\boldsymbol{\mu})
\widetilde Q_\Ups(\boldsymbol{\mu})
=
\sum_{\boldsymbol{\alpha}}
C_{\mathsf{C}}(\boldsymbol{\alpha})
Q_\Ups(\boldsymbol{\alpha})
=
\Tr(\mathsf{C}\Ups_{\cA}).
\label{eq:frame-invariant-pairing}
\end{equation}
In either collection of frames, the admissible phase-space couplings
are precisely the coordinate images of
$\mathfrak{C}(\rho_1,\ldots,\rho_N)$.
The frame transformation therefore gives a bijection between
the two admissible sets and preserves the objective function. Hence the optimal values coincide.
\end{proof}

For a fixed operator $\mathsf{C}$, changing the local frames changes
only its coordinates and those of the couplings. By contrast,
keeping cost function fixed while changing the frames
generally changes the associated cost operator and hence the
transport problem.

\subsection{Quantum Kantorovich duality}

We now establish the Kantorovich duality for spatial quantum optimal transport and discuss how to represent it in phase space by generalizing the discussing in \cite{chakrabarti2019quantum,Cole2023On} for positive semidefinite cost operator to general case.

The key observation is that, for Hermitian cost operator, if we consider the objective function $\Tr \sC \Ups$ (without modulus), the Kantorovich duality is actually the dual problem of SDP of  quantum optimal transport.
We consider the primal problem of an SDP of the following form \cite{vandenberghe1996semidefinite,watrous2009semidefinite,watrous2012simpler}:
\begin{equation}
    \begin{aligned}
        \inf_{X} \quad & \langle C,X\rangle =  \operatorname{Tr}(C X) \\
        \text{subject to} \quad & X \ge 0, \\
        & \mathcal{E}(X) = B,
    \end{aligned}
\end{equation}
where \(\mathcal{E} : \Herm(\mathcal{H}) \to \Herm(\mathcal{V})\) is linear and Hermiticity preserving. The dual problem is
\begin{equation}
    \begin{aligned}
        \sup_{Y} \quad & \langle B, Y \rangle =\Tr (BY) \\
        \text{subject to} \quad & \mathcal{E}^{\dagger}(Y) \le C,
    \end{aligned}
\end{equation}
where the inner product is the Hilbert--Schmidt inner product and \(\mathcal{E}^{\dagger}\) is the adjoint of \(\mathcal{E}\) with respect to this inner product.

Assume that the cost operator is Hermitian (not necessarily positive semidefinite), \(\sC = \sC^{\dagger}\). Then spatial optimal transport can be recast as an SDP
\begin{equation}
\begin{split}
\mathrm{QMK}_{\mathsf{C}}(\rho_A,\rho_B)
=
\inf_{\Ups}\quad&
\Tr(\mathsf{C}\Ups)\\
\text{subject to}\quad&
\Ups\geq0,\\
&
\Tr_B\Ups=\rho_A,\\
&
\Tr_A\Ups=\rho_B.
\end{split}
\label{eq:spatial-primal-SDP}
\end{equation}
The marginal constraints imply $\Tr\Ups=1$.Thus the feasible
set is nonempty and compact, and the minimum is attained. In
particular, the eigenvalues of $\sC$ give the following bounds
\begin{equation}
\lambda_{\min}(\mathsf{C})
\leq
\mathrm{QMK}_{\mathsf{C}}(\rho_A,\rho_B)
\leq
\lambda_{\max}(\mathsf{C}),
\end{equation}
since $\lambda_{\min}(\mathsf{C})
\leq
\Tr \sC\Ups
\leq
\lambda_{\max}(\mathsf{C})$
for every density operator $\Ups$.

Notice that for SDP in Eq.~\eqref{eq:spatial-primal-SDP}
\[
\begin{aligned}
\mathcal E:\Herm(\mathcal H_A\otimes\mathcal H_B)
&\longrightarrow
\Herm(\mathcal H_A)\oplus\Herm(\mathcal H_B),\\
X&\longmapsto(\Tr_BX,\Tr_AX).
\end{aligned}
\]
the adjoint is given by
adjoint is
\[
\begin{aligned}
\mathcal E^{\dagger}:
\Herm(\mathcal H_A)\oplus\Herm(\mathcal H_B)
&\longrightarrow\Herm(\mathcal H_A\otimes\mathcal H_B),\\
(Y_A,Y_B)&\longmapsto
Y_A\otimes I_B+I_A\otimes Y_B.
\end{aligned}
\]
the dual problem in this setting is called quantum Kantorovich duality:
\begin{equation}
\widehat{\mathrm{QMK}}_{\mathsf{C}}(\rho_A,\rho_B)
=
\sup_{X=X^\dagger,\,Y=Y^\dagger}
\left\{
\Tr(X\rho_A)+\Tr(Y\rho_B):
X\otimes I_B+I_A\otimes Y\leq\mathsf{C}
\right\}.
\label{eq:spatial-Kantorovich-dual}
\end{equation}
where $X,Y$ can be regarded as quantum generalization of Kantorovich potential.

\begin{theorem}[Quantum Kantorovich duality for Hermitian cost operator]
\label{thm:spatial-Kantorovich-duality}
For every Hermitian cost operator $\mathsf{C}$, quantum Kantorovich duality is given in Eq.~\eqref{eq:spatial-Kantorovich-dual}.
The weak duality of SDP implies that $\widehat{\mathrm{QMK}}_{\mathsf{C}}(\rho_A,\rho_B) \le \mathrm{QMK}_{\mathsf{C}}(\rho_A,\rho_B)$.
Since the primal is feasible and the dual is strictly feasible, so Slater’s theorem applies (see Theorem~\ref{thm:SDPslater} in Appendix), we in fact have strong duality:
\begin{equation}\label{eq:strongDual}
    \widehat{\mathrm{QMK}}_{\mathsf{C}}(\rho_A,\rho_B) =
    \mathrm{QMK}_{\mathsf{C}}(\rho_A,\rho_B)
\end{equation}
The primal infimum is always attained. 
If $\rho_A$ and $\rho_B$ are strictly positive, the dual supremum is also attained.
For arbitrary marginals, both optima are attained after restricting
the Hilbert spaces to their supports and compressing the cost
operator to the resulting tensor-product space.
\end{theorem}

\begin{proof}
Introduce Hermitian Lagrange multipliers $X$ and $Y$ for the
marginal constraints. The Lagrangian is
\begin{align}
L(\Ups,X,Y)
={}&
\Tr(\mathsf{C}\Ups)
+\Tr\left[X(\rho_A-\Tr_B\Ups)\right]
+\Tr\left[Y(\rho_B-\Tr_A\Ups)\right]
\nonumber\\
={}&
\Tr(X\rho_A)+\Tr(Y\rho_B)
\nonumber\\
&+
\Tr\left[
\left(
\mathsf{C}-X\otimes I_B-I_A\otimes Y
\right)\Ups
\right].
\label{eq:spatial-Lagrangian}
\end{align}
The infimum over $\Ups\geq0$ is finite precisely when
\begin{equation}
\mathsf{C}-X\otimes I_B-I_A\otimes Y\geq0.
\label{eq:dual-operator-inequality}
\end{equation}
When the operator is positive, the infimum of the last
term is zero. This gives the stated dual problem.

By Slater's condition \cite[Theorem~1.1]{watrous2012simpler}, since 
the primal feasible set is non-empty and for every Hermitian $\mathsf{C}$:
choose $X=-tI_A$ and $Y=-tI_B$, with
$2t>-\lambda_{\min}(\mathsf{C})$. Then
\begin{equation}
\mathsf{C}-X\otimes I_B-I_A\otimes Y
=
\mathsf{C}+2tI_A\otimes I_B>0.
\end{equation}
Then we have strong duality as in Eq.~\eqref{eq:strongDual} and the there exists a primal feasible spatial state $\Ups_{\star}$ for which $\mathrm{QMK}_{\mathsf{C}}(\rho_A,\rho_B)=\Tr [\sC \Ups_{\star}]$.

Similarly, when $\rho_A$ amd $\rho_B$ are strictly positive, since dual feasible set is non-empty and there exists $\rho_A\otimes \rho_B>0$ is primally feasible, we have strong duality as in Eq.~\eqref{eq:strongDual} and there exists $X_{\star}$ and $Y_{\star}$ such that $\Tr[X_{\star}\rho_A]+\Tr[Y_{\star}\rho_B]=\widehat{\mathrm{QMK}}_{\mathsf{C}}(\rho_A,\rho_B)$.

The above argument directly implies that, for general $\rho_A$ and $\rho_B$ when restricted to tensor product space of their supports, both optima are attained.
\end{proof}

The dual variables also possess phase-space representations. Define
\begin{equation}
\varphi_X(\alpha)=\Tr(D_\alpha X),
\quad
\psi_Y(\beta)=\Tr(E_\beta Y).
\label{eq:KD-Kantorovich-potentials}
\end{equation}
Then
\begin{equation}
\Tr(X\rho_A)+\Tr(Y\rho_B)
=
\sum_\alpha
\varphi_X(\alpha)p_{\rho_A}(\alpha)
+
\sum_\beta
\psi_Y(\beta)p_{\rho_B}(\beta).
\label{eq:KD-dual-objective}
\end{equation}
These functions are the quantum analogues of Kantorovich potentials in phase space.
Unlike classical Kantorovich potentials, however, they are constrained
by the operator inequality
\eqref{eq:dual-operator-inequality}, rather than only by pointwise
inequalities.

For general non-Hermitian cost operator $\sC$, we need consider the objective function with modulus $|\Tr \sC \Ups|$, in this case, we have the following duality:

\begin{theorem}[Quantum Kantorovich duality for non-Hermitian cost operator]
\label{thm:complex-KD-duality}
For general cost operator $\mathsf{C}\in \mathbf{B}(\cH)$, the Kantorovich duality is given by
\begin{equation} \label{eq:complex-spatial-Kantorovich-duality}
  \widehat{\mathrm{QKD}}_{\mathsf{C}}(\rho_A,\rho_B) =  \sup_{\substack{u_1,u_2\in\mathbb{R}\\u_1^2+u_2^2\leq1}}
\;
\sup_{X=X^\dagger,\,Y=Y^\dagger}
\left\{
\Tr(X\rho_A)+\Tr(Y\rho_B):
X\otimes I_B+I_A\otimes Y
\leq
u_1\mathsf{C}_{\mathrm{R}}
+
u_2\mathsf{C}_{\mathrm{I}}
\right\}.
\end{equation}
The strong duality holds as in Eq.~\eqref{eq:strongDual}.
The primal infimum is always attained. 
If $\rho_A$ and $\rho_B$ are strictly positive, the dual supremum is also attained.
For arbitrary marginals, both optima are attained after restricting
the Hilbert spaces to their supports and compressing the cost
operator to the resulting tensor-product space.
\end{theorem}

\begin{proof}
For every complex number $z$, when treated as two-dimensional real vectors, Cauchy–Schwarz inequality implies
\begin{equation}
|z|
=
\sup_{u_1^2+u_2^2\leq1}
\left(
u_1\mathrm{Re}\,z+
u_2\mathrm{Im}\,z
\right)
=\sup_{|u|\leq 1} \operatorname{Re} \Bar{u}z.
\label{eq:modulus-support-function}
\end{equation}
Consequently,
\begin{align}
&
\inf_{\Ups\in\mathfrak{C}(\rho_A,\rho_B)}
\left|\Tr(\mathsf{C}\Ups)\right|
\nonumber\\
&\quad=
\inf_{\Ups\in\mathfrak{C}(\rho_A,\rho_B)}
\sup_{u_1^2+u_2^2\leq1}
\Tr\left[
(u_1\mathsf{C}_{\mathrm{R}}
+u_2\mathsf{C}_{\mathrm{I}})\Ups
\right].
\label{eq:complex-minmax-first}
\end{align}
The spatial coupling set $\mathfrak{C}(\rho_A,\rho_B)$ and the closed unit disk are nonempty, compact,
and convex, and the objective is real-valued, continuous, and
bilinear. The minimax theorem (see Theorem~\ref{thm:minimax} in Appendix) therefore gives
\begin{align}
&
\inf_{\Ups\in\mathfrak{C}(\rho_A,\rho_B)}
\left|\Tr(\mathsf{C}\Ups)\right|
\nonumber\\
&\quad=
\sup_{u_1^2+u_2^2\leq1}
\inf_{\Ups\in\mathfrak{C}(\rho_A,\rho_B)}
\Tr\left[
(u_1\mathsf{C}_{\mathrm{R}}
+u_2\mathsf{C}_{\mathrm{I}})\Ups
\right].
\label{eq:complex-minmax-second}
\end{align}
For each fixed $(u_1,u_2)$, the operator
$u_1\mathsf{C}_{\mathrm{R}}+u_2\mathsf{C}_{\mathrm{I}}$
is Hermitian. Applying
Theorem~\ref{thm:spatial-Kantorovich-duality} to the inner problem
proves \eqref{eq:complex-spatial-Kantorovich-duality}.

The primal infimum is attained by compactness and continuity.
Moreover, the inner optimal value in
\eqref{eq:complex-minmax-second} is a continuous function of
$(u_1,u_2)$, so its supremum over the closed unit disk is attained.
On the support space of $\rho_A,\rho_B$, Theorem~\ref{thm:spatial-Kantorovich-duality}
guarantees dual attainment for the corresponding Hermitian cost.
Since unit disk is compact, the joint dual supremum is attained on that space.
\end{proof}

Note that for the real-part objective function \(\operatorname{Re} \Tr [\sC \Ups] =\Tr [\sC_{\rm R}\Ups]\), it reduces to the standard setting by replacing the cost operator with \(\sC_{\rm R}\). Theorem~\ref{thm:spatial-Kantorovich-duality} can thus be applied to give the Kantorovich duality.

The complex-cost problem has the conic formulation
\begin{equation}
\begin{split}
\inf_{\Ups,t}\quad &t\\
\text{subject to}\quad&
\Ups\geq0,\quad
\Tr_B\Ups=\rho_A,\quad
\Tr_A\Ups=\rho_B,\\
&
\sqrt{
\Tr(\mathsf{C}_{\mathrm{R}}\Ups)^2+
\Tr(\mathsf{C}_{\mathrm{I}}\Ups)^2
}
\leq t.
\end{split}
\label{eq:complex-cost-conic-primal}
\end{equation}
The last constraint is a three-dimensional second-order cone
constraint. Equivalently, writing
$a=\Tr(\mathsf{C}_{\mathrm{R}}\Ups)$ and
$b=\Tr(\mathsf{C}_{\mathrm{I}}\Ups)$, it is the linear matrix
inequality
\begin{equation}
\begin{pmatrix}
t+a & b\\
b & t-a
\end{pmatrix}
\geq0.
\label{eq:complex-cost-LMI}
\end{equation}
Hence the complex-cost problem can also be expressed entirely as
a semidefinite program.

For a Hermitian but not necessarily positive cost $\mathsf{C}$,
define
\begin{equation}
m_{\mathsf{C}}
=
\inf_{\Ups\in\mathfrak{C}(\rho_A,\rho_B)}
\Tr(\mathsf{C}\Ups),
\quad
M_{\mathsf{C}}
=
\sup_{\Ups\in\mathfrak{C}(\rho_A,\rho_B)}
\Tr(\mathsf{C}\Ups).
\label{eq:attainable-cost-interval}
\end{equation}
Compactness and convexity of the coupling set imply that the
attainable expectation values form the closed interval
$[m_{\mathsf{C}},M_{\mathsf{C}}]$. Therefore,
\begin{equation}
\inf_{\Ups\in\mathfrak{C}(\rho_A,\rho_B)}
\left|\Tr(\mathsf{C}\Ups)\right|
=
\begin{cases}
m_{\mathsf{C}},
&
m_{\mathsf{C}}>0,\\[1mm]
0,
&
m_{\mathsf{C}}\leq0\leq M_{\mathsf{C}},\\[1mm]
-M_{\mathsf{C}},
&
M_{\mathsf{C}}<0.
\end{cases}
\label{eq:Hermitian-modulus-solution}
\end{equation}
Thus the modulus problem for a Hermitian cost can be solved by
two ordinary quantum Kantorovich semidefinite programs, with
cost operators $\mathsf{C}$ and $-\mathsf{C}$, respectively.

\section{Temporal optimal transport in temporal phase space}
\label{sec:temporal-optimal-transport}

We now apply our framework to optimal transport in temporal phase space. 
The essential difference from the spatial problem is that a temporal coupling is not, in general, a positive semidefinite density operator. Instead, it is a generally non-Hermitian and non-normal temporal state obtained by temporally linking an initial state with the Jamio{\l}kowski operators of the intervening quantum evolutions. Its Kirkwood-Dirac representation is consequently a generally complex quasiprobability distribution.

Let $t_0<t_1<\cdots<t_n$ be a discrete set of times, and let $\rho_{t_k}\in\mathbf{State}(\mathcal{H}_{t_k})$ be prescribed single-time
density operators. We write
$\boldsymbol{\rho}
=
(\rho_{t_0},\rho_{t_1},\ldots,\rho_{t_n})$.
At each time $t_k$, let
$\{F^{(k)}_{\alpha_k}\}$ be a minimal
IC-POVM frame, with dual frame
$\{D^{(k)}_{\alpha_k}\}_{\alpha_k\in I_k}$.
For an operator $\Upsilon$ on
$\mathcal{H}_{t_n}\otimes\cdots\otimes\mathcal{H}_{t_0}$, its temporal
Kirkwood--Dirac representation is
\begin{equation}
Q_{\Upsilon}(\alpha_n,\ldots,\alpha_0)
=\Tr[F_{\bfalpha}\Ups]=
\Tr\left[
\left(
F^{(n)}_{\alpha_n}\otimes\cdots\otimes
F^{(0)}_{\alpha_0}
\right)\Upsilon
\right].
\label{eq:temporal-transport-KD-representation}
\end{equation}
The inverse relation reads
\begin{equation}
\Upsilon
=\sum_{\bfalpha} Q_{\Upsilon}(\bfalpha)D_{\bfalpha}
=
\sum_{\alpha_n,\ldots,\alpha_0}
Q_{\Upsilon}(\alpha_n,\ldots,\alpha_0)
D^{(n)}_{\alpha_n}\otimes\cdots\otimes
D^{(0)}_{\alpha_0}.
\label{eq:temporal-transport-reconstruction}
\end{equation}
Thus a temporal Kirkwood-Dirac coupling and its temporal-state representation contain
exactly the same information.

For a temporal phase-space cost
$C(\alpha_n,\ldots,\alpha_0)$, define the corresponding
cost operator by
\begin{equation}
\mathsf{C}
=\sum_{\bfalpha} C(\bfalpha)F_{\bfalpha}=
\sum_{\alpha_n,\ldots,\alpha_0}
C(\alpha_n,\ldots,\alpha_0)
F^{(n)}_{\alpha_n}\otimes\cdots\otimes
F^{(0)}_{\alpha_0}.
\label{eq:temporal-cost-operator}
\end{equation}
Then
\begin{equation}
\sum_{\alpha_n,\ldots,\alpha_0}
C(\alpha_n,\ldots,\alpha_0)
Q_{\Upsilon}(\alpha_n,\ldots,\alpha_0)
=
\Tr(\mathsf{C}\Upsilon).
\label{eq:temporal-cost-pairing}
\end{equation}
The temporal quantum optimal transport  can therefore be formulated
either in temporal phase space or directly in terms of temporal states.

\subsection{Fixed-branch case}
\label{subsec:fixed-branch-temporal-transport}

We first consider the fixed-branch temporal state \cite{jia2026unifiedspatiotemporalquantumstates}, where the temporal link product is given in a fixed order. As we will see, in two-time case, this is equivalent to quantum channel optimal transport (see \cite{DePalma2021QuantumMKchannal} for the discussion of quantum channel optimal transport).
Consider the branch
\begin{equation}
\mathtt{B}
=
(t_n^{s_n},\ldots,t_0^{s_0}),
\quad
s_k\in\{L,R\}.
\label{eq:fixed-temporal-branch}
\end{equation}
The branch determines the ordering of the temporal link products and,
equivalently, which copy of each doubled temporal slot remains
uncontracted. For the temporal Kirkwood--Dirac distribution, this determines the position where we insert the IC-POVMs, e.g., \(Q(\alpha_n,\cdots,\alpha_0)=\Tr[F_{\alpha_n}^{(n)} \cE_{n} (\cdots \cE_1(F_{\alpha_0}^{(0)} \rho_{t_0} ) F_{\alpha_1}^{(1)}  )]\) for which \(\mathtt{B}= (t_n^L,\cdots, t_1^R, t_0^L)\).

For every $k=1,\ldots,n$, let $\cE_k:\mathbf{B}(\cH_{t_{k-1}}) \to \mathbf{B}(\cH_{t_{k}})$
be quantum channels, its Choi operator is defined by
\begin{equation}
J'_k=C[\cE_k]=d_{t_{k-1}}\cE\otimes \operatorname{id} (|\rm GHZ\rangle \langle |\rm GHZ| )
\end{equation}
where $|\mathrm{GHZ}\rangle=\frac{1}{\sqrt{d_{t_{k-1}}}}\sum_{i=1}^{d_{t_{k-1}}} |ii\rangle$ is  Greenberger-Horne-Zeilinger (GHZ) state. It's clear that Choi operator is positive semidefinite and $\Tr_{t_k}C[\cE_k]=I_{t_{k-1}}$.
The Jamio{\l}kowski operator is partial transpose of Choi operator
\begin{equation}
J_k
=
J[\mathcal{E}_k]
=
\sum_{i,j}
\mathcal{E}_k(|i\rangle\langle j|)
\otimes
|j\rangle\langle i|=C[\cE_k]^{T_{t_{k-1}}}.
\label{eq:temporal-transport-Jamiolkowski}
\end{equation}
This convention is adapted to ensure $\Tr_{t_{k-1}}[(A\otimes B)J_k]
= 
A\mathcal{E}_k(B)$.

In Ref.~\cite{jia2026unifiedspatiotemporalquantumstates}, we show that the temporal state for branch $\mathtt{B}$ is given by temporal link product
\begin{equation}
    \Ups^{\ttB}_{t_n,\cdots,t_0}=J_n \star_{\ttB} J_{n-1}  \star_{\ttB} \cdots  \star_{\ttB} J_1  \star_{\ttB} \rho_{t_0}.
\end{equation}
At time $t_k$, the reduced state $\Tr_{\widehat{t}_k}  \Ups^{\ttB}_{t_n,\cdots,t_0} =\rho_{t_k}=\mathcal{E}_k\circ \cdots \circ \cE_1(\rho_{t_0})$.
We denote the set of all fixed-branch temporal state with marginals $\rho_{t_0}$, $\cdots$, $\rho_{t_n}$ as $\mathfrak{C}^{\ttB}(\rho_{t_0},\cdots,\rho_{t_n})$. 
The condition that $\mathcal{E}_k$ transports the prescribed state
$\rho_{t_{k-1}}$ to $\rho_{t_k}$ is
\begin{equation}
\mathcal{E}_k(\rho_{t_{k-1}})
=
\rho_{t_k}.
\label{eq:state-transport-constraint}
\end{equation}
In terms of Jamio{\l}kowski operator $J_k$, this becomes
\begin{equation}
\Tr_{t_{k-1}}
\left[
(I_{t_k}\otimes\rho_{t_{k-1}})J_k
\right]
=
\rho_{t_k},
\label{eq:Jamiolkowski-state-transport}
\end{equation}
Define the set of admissible one-step transport Choi operator by
\begin{equation}
\begin{split}
\mathbf{Choi}_k
(\rho_{t_{k-1}}\rightarrow\rho_{t_k})
=
\Bigl\{
J_k:\;&
J_k\geq0,\quad
\Tr_{t_k}{J}_k=I_{t_{k-1}},\\
&
\Tr_{t_{k-1}}
\left[
{J}_k
(I_{t_k}\otimes\rho_{t_{k-1}}^{T})
\right]
=
\rho_{t_k}
\Bigr\}.
\end{split}
\label{eq:one-step-transport-Choi-set}
\end{equation}
This set is a compact convex set. It is always nonempty,
since the replacement channel
\begin{equation}
\mathcal{E}_k(X)
=
\Tr(X)\rho_{t_k}
\label{eq:replacement-channel}
\end{equation}
belongs to it.
The set of admissible 
Jamio{\l}kowski operator
is 
\begin{equation}
    \mathbf{Jami}_k
(\rho_{t_{k-1}}\rightarrow\rho_{t_k})=\{J: J^{T{t_{k-1}}} \in \mathbf{Choi}_k
(\rho_{t_{k-1}}\rightarrow\rho_{t_k})\}
\end{equation}
The fixed-branch temporal-state
coupling set is thus of the form
\begin{equation}
\begin{split}
\mathfrak{C}^{\ttB}(\rho_{t_0},\cdots,\rho_{t_n})
=
\Bigl\{
&
\Upsilon^{\mathtt{B}}
(\rho_{t_0};
J_1,\cdots,J_n)
:\nonumber\\
&
J_k\in
 \mathbf{Jami}_k
(\rho_{t_{k-1}}\rightarrow\rho_{t_k}),
\quad k=1,\ldots,n
\Bigr\}.
\end{split}
\label{eq:fixed-branch-temporal-state-set}
\end{equation}
Each element of this set satisfies
\begin{equation}
\Tr_{t_n,\ldots,\widehat{t_k},\ldots,t_0}
\Upsilon^{\mathtt{B}}
=
\rho_{t_k},
\quad
k=0,\ldots,n.
\label{eq:fixed-branch-single-time-marginals}
\end{equation}
Fix a minimal IC-POVM frame, The corresponding set of fixed-branch temporal Kirkwood-Dirac coupling is
\begin{equation}
\mathbb{KD}^{\mathtt{B}}
(Q_{t_0},\cdots,Q_{t_n})
=
\left\{
Q_{\Upsilon}:
\Upsilon\in
\mathfrak{C}^{\mathtt{B}}(\rho_{t_0},\cdots,\rho_{t_n})
\right\}.
\label{eq:fixed-branch-KD-plan-set}
\end{equation}
 the map
\begin{equation}
\Upsilon\longmapsto Q_{\Upsilon}
\end{equation}
is a bijection from
$\mathfrak{C}^{\mathtt{B}}(\boldsymbol{\rho})$ onto
$\mathbb{KD}^{\mathtt{B}}(\boldsymbol{\rho})$. Its inverse is given by
\eqref{eq:temporal-transport-reconstruction}.

The fixed-branch temporal Monge-Kantorovich optimal transport problem is defined as 
\begin{equation}
\begin{split}
\operatorname{QMK}_{\mathtt{B},C}
(\boldsymbol{\rho})
&:=
\inf_{Q\in
\mathbb{KD}^{\mathtt{B}}
(Q_{t_0},\cdots,Q_{t_n})}
\left|
\sum_{\alpha_n,\ldots,\alpha_0}
C(\alpha_n,\ldots,\alpha_0)
Q(\alpha_n,\ldots,\alpha_0)
\right|
\nonumber\\
&=
\inf_{\Upsilon\in
\mathfrak{C}^{\mathtt{B}}(\boldsymbol{\rho})}
\left|
\Tr(\mathsf{C}\Upsilon)
\right|=:\operatorname{QMK}_{\mathtt{B},\sC}
(\boldsymbol{\rho}).
\end{split}
\label{eq:fixed-branch-temporal-MK}
\end{equation}
Equivalently,
\begin{equation}
\begin{split}
\operatorname{QMK}_{\mathtt{B},\sC}
(\boldsymbol{\rho})
=
\inf_{\substack{
{J}_k\in
\mathbf{Jami}_k
(\rho_{t_{k-1}}\rightarrow\rho_{t_k})\\
k=1,\ldots,n}}
\left|
\Tr\left[
\mathsf{C}\,
\Upsilon^{\mathtt{B}}
\left(
\rho_{t_0};
{J}_1,
\ldots,
{J}_n
\right)
\right]
\right|.
\end{split}
\label{eq:fixed-branch-Choi-optimization}
\end{equation}
Note that $\rho_{t_0}$ is one of the local marginals, thus it is fixed fixed-branch temporal optimal transport, we can only vary the quantum channels. In this sense, the problem is equivalent to the quantum channel optimal transport. However, the formalism in \cite{DePalma2021QuantumMKchannal} can only work in two-time case, our formalism gives a natural multi-marginal generalization.
The quantum channel optimal transport in  \cite{DePalma2021QuantumMKchannal} in fact considers temporal couplings of the form $\Xi= \cE\otimes \id (|\sqrt{\rho_{t_0}}\rrangle\llangle \sqrt{\rho_{t_0}}|)$, where $\cE$ is required to give a fixed output $\cE(\rho_{t_0})=\rho_{t_1}$. Let $\mathcal{F}_{\cE,\rho_{t_0}}=\cE(\sqrt{\rho_{t_0}}\bullet \sqrt{\rho_{t_0}})$,
using Eq.~\eqref{eq:vectorization} in Appendix, we see that $\Xi$ is indeed the Choi operator $\Xi=C[\mathcal{F}_{\cE,\rho_{t_0}}]$, from which it is clear that $\Tr_{t_0}\Xi=\rho_{t_1}$. It can also be expressed as $\Xi=\cE\otimes \id ((\mathds{1}\otimes \sqrt{\rho}^T)|\Omega\rrangle\llangle \Omega|(\mathds{1}\otimes \sqrt{\rho}^T))$, from which we have $\Tr_{t_1} \Xi=\rho_{t_0}^T$. For $t_0$, there is a transpose, which makes the marginals for the two times not on an equal footing.
In our setting via setting $\Ups=J[\cE](\mathds{1}\otimes \rho_{t_0})$, which gives $\Tr_{t_0}\Ups=\rho_{t_1}$ and $\Tr_{t_1}\Ups=\rho_{t_0}$, which is more symmetric. In two-time case, they are essentially equivalent. 
However, the expression for $\Xi$ in \cite{DePalma2021QuantumMKchannal} cannot be generalized straightforwardly to the multi-marginal setting, our setting give a more natural definition of temporal quantum optimal transport. For instance, for three-time with $\ttB=(t_2^R,t_1^R,t_0^R)$, we have $\Ups=J[\cE_2]J[\cE_1](\mathds{1}\otimes \rho_{t_0})$, which gives correct marginals $\rho_{t_0}$, $\rho_{t_1}$ and $\rho_{t_2}$. A more detailed discussion will be given in a forthcoming work~\cite{kang2026quantum}.

\begin{proposition}[Existence and multiaffine structure]
\label{prop:fixed-branch-multiaffine}
The infimum of temporal optimal transport in Eq.~\eqref{eq:fixed-branch-Choi-optimization} is attained. The map
\begin{equation}
({J}_1,
\cdots,
{J}_n)
\longmapsto
\Upsilon^{\mathtt{B}}
\left(
\rho_{t_0};
{J}_1,
\cdots,
{J}_n
\right)
\label{eq:multiaffine-temporal-state-map}
\end{equation}
is linear in each ${J}_k$ when the remaining Jamio{\l}kowski operators are
held fixed. It is not, in general, jointly affine\footnote{By ``affine'' we mean preserving convex mixtures.} when $n\geq2$.
Consequently, the fixed-branch temporal quantum optimal transport problem is a multi-convex optimization problem, but not generally a convex optimization problem.
\end{proposition}

\begin{proof}
Each set $ \mathbf{Jami}_k
(\rho_{t_{k-1}}\rightarrow\rho_{t_k})$ is compact. Their Cartesian product is therefore compact. Fixed-branch temporal
link products $\star_{\ttB}$ is continuous, thus the objective function in \eqref{eq:fixed-branch-Choi-optimization} is continuous. The infimum is therefore attained.

Every Jamio{\l}kowski operator occurs exactly once in the temporal link-product expression. Holding all other factors fixed makes the temporal state linear in that Jamio{\l}kowski operator. When two or more Jamio{\l}kowski operators are varied simultaneously, products such as ${J}_{k+1}{J}_k$ occur. The resulting map is therefore multiaffine rather than jointly affine.
\end{proof}

This distinction of two-time ($n=1$) and multi-time ($n \geq 2$) is important. For two times, there is only one
variable Choi/Jamio{\l}kowski operator, and the fixed-branch problem is convex. For
three or more times, independent optimization over all intervening
channels generally produces a nonconvex factorized-process problem.

\subsubsection{Quantum Kantorovich duality of the two-time temporal quantum optimal transport}

Consider two-time case $t_0<t_1$. There are only two independent branches\footnote{For final step, choosing $L$ and $R$ are equivalent,  see Ref.~\cite{jia2026unifiedspatiotemporalquantumstates} for details.} $\ttB=(t_1^L,t_0^L)$ and $\ttB^c=(t_1^R,t_0^R)$ which gives two temporal states
\begin{equation}
    \Ups^{\ttB} =(\mathds{1}\otimes \rho_{t_0})J_1, \quad  \Ups^{\ttB^c} =  J_1(\mathds{1}\otimes \rho_{t_0}).
\end{equation}
Set $A_{\mathtt{B},\mathsf{C}}=\sC(\mathds{1}\otimes \rho_{t_0})$, we have
\begin{equation}
\Tr\left[
\mathsf{C}\,
\Upsilon^{\mathtt{B}}
\left(
\rho_{t_0};
J_1
\right)
\right]
=
\Tr(A_{\mathtt{B},\mathsf{C}}J_1)
\label{eq:effective-two-time-cost}
\end{equation}
for every $J_1$. Thus the temporal optimal transport becomes a optimization over all $J_1\in  \mathbf{Jami}
(\rho_{t_{0}}\rightarrow\rho_{t_1})$ with cost operator $A_{\mathtt{B},\mathsf{C}}$.

To establish the quantum Kantorovich duality, it is more convenient to work with Choi operator. We can use Lemma~\ref{lemma:PartialTranspose} in Appendix to transform the temporal optimal transport problem into
\begin{align}
&
\inf_{J'\in
\mathbf{Choi}
(\rho_{t_0}\rightarrow\rho_{t_1})}
\left|
\Tr(A_{\mathtt{B},\mathsf{C}}^{T_{t_0}}{J}')
\right|
\end{align}
In this case, the optimization is taken over positive semidefinite operators.

\begin{theorem}[Quantum Kantorovich duality of two-time fixed-branch temporal quantum optimal transport]
\label{thm:two-time-temporal-transport-SDP}
For $u\in\mathbb{C}$, define the Hermitian operator
\begin{equation}
H_{\mathtt{B},\mathsf{C}}(u)
=
\frac{1}{2}
\left(
\overline{u} A_{\mathtt{B},\mathsf{C}}^{T_{t_0}}
+
u (A_{\mathtt{B},\mathbb{C}}^{T_{t_0}})^\dagger
\right) \in \mathbf{Herm}(\cH_{t_1}\otimes \cH_{t_0}).
\label{eq:phase-rotated-effective-cost}
\end{equation}
The quantum Kantorovich duality of temporal quantum optimal transport is given by
\begin{equation}
\begin{split}
\widehat{\operatorname{QMK}}_{\mathtt{B},\sC}
(\rho_{t_0},\rho_{t_1})
=
\sup_{\substack{u\in\mathbb{C},\ |u|\leq1\\
X=X^\dagger,\ Y=Y^\dagger}}
\quad&
\Tr(X)+\Tr(Y\rho_{t_1})\\
\text{subject to}\quad&
I_{t_1}\otimes X
+
Y\otimes\rho_{t_0}^{T}
\leq
H_{\mathtt{B},\mathsf{C}}(u).
\end{split}
\label{eq:two-time-temporal-transport-dual}
\end{equation}
Here $X$ acts on $\mathcal{H}_{t_0}$ and $Y$ acts on
$\mathcal{H}_{t_1}$. We have 
\begin{equation}\label{eq:temporalKantDual}
   {\operatorname{QMK}}_{\mathtt{B},\sC}
(\rho_{t_0},\rho_{t_1}) =\widehat{\operatorname{QMK}}_{\mathtt{B},\sC}
(\rho_{t_0},\rho_{t_1})
\end{equation}
The primal infimum is always attained. If $\rho_{t_1}$ is full rank, the dual supremum is always attained.
\end{theorem}

\begin{proof}
For every $z\in\mathbb{C}$,
\begin{equation}
|z|
=
\sup_{|u|\leq1}
\operatorname{Re}(\overline{u}z).
\label{eq:complex-modulus-support}
\end{equation}
This implies
\begin{align}
&
\inf_{J'\in
\mathbf{Choi}
(\rho_{t_0}\rightarrow\rho_{t_1})}
\left|
\Tr(A_{\mathtt{B},\mathsf{C}}^{T_{t_0}}{J}')
\right|
\nonumber\\
&\quad=
\inf_{J'\in
\mathbf{Choi}
(\rho_{t_0}\rightarrow\rho_{t_1})}
\sup_{|u|\leq1}
\Tr\left[
H_{\mathtt{B},\mathsf{C}}(u){J}'
\right].
\label{eq:two-time-minmax}
\end{align}
The $\mathbf{Choi}
(\rho_{t_0}\rightarrow\rho_{t_1})$ and the unit disk are compact and convex, and the
objective is bilinear. The minimax theorem therefore gives
\begin{equation}
\sup_{|u|\leq1}
 \inf_{J'\in
\mathbf{Choi}
(\rho_{t_0}\rightarrow\rho_{t_1})}
\Tr\left[
H_{\mathtt{B},\mathsf{C}}(u){J}
\right].
\label{eq:two-time-minmax-interchanged}
\end{equation}
For fixed $u$, the inner problem of Eq.~\eqref{eq:two-time-minmax-interchanged} is a SDP
\begin{equation}
\begin{split}
\inf_{J'\in
\mathbf{Choi}
(\rho_{t_0}\rightarrow\rho_{t_1})} &
\Tr\left[
H_{\mathtt{B},\mathsf{C}}(u) {J}'
\right]\\
\text{subject to}\quad&
{J}'\ge 0,\\
&
\Tr_{t_1}{J}'=I_{t_0},\\
&
\Tr_{t_0}
\left[
{J}'
(I_{t_1}\otimes\rho_{t_0}^{T})
\right]
=
\rho_{t_1}.
\end{split}
\label{eq:two-time-fixed-phase-primal}
\end{equation}
Introducing Hermitian multipliers $X$ and $Y$ for the last two
equalities gives the dual
\begin{equation}
\begin{split}
\sup_{X=X^\dagger,\,Y=Y^\dagger}\quad&
\Tr(X)+\Tr(Y\rho_{t_1})\\
\text{subject to}\quad&
I_{t_1}\otimes X
+
Y\otimes\rho_{t_0}^{T}
\leq
H_{\mathtt{B},\mathsf{C}}(u).
\end{split}
\label{eq:two-time-fixed-phase-dual}
\end{equation}

Since primal feasible set is non-empty since replacement channel gives a feasible Choi operator, and we define
\begin{equation}
    \cE: J'\mapsto (\Tr_{t_1} J', \Tr J' \rho_{t_0}^{T})
\end{equation}
the adjoint is 
\begin{equation}
    \cE^{\dagger}:(X,Y) \mapsto I_{t_1}\otimes X +Y\otimes \rho_{t_0}^T.
\end{equation}
Set $X=-tI_{t_0}$ and $Y=0$, $\cE^{\dagger}(X,Y)=-t I_{t_1t_0}$, when $t> -\lambda_{\rm min} (H_{\ttB,\sC}(u))$, we have $\cE^{\dagger}(X,Y) < H_{\ttB,\sC}(u)$, they satisfy Slater’s condition, thus the primal infimum is always attained  and Eq.~\eqref{eq:temporalKantDual} holds.

If $\rho_{t_1}$ is full rank, the replacement channel has the strictly
positive Choi matrix
\begin{equation}
{J}'=\rho_{t_1}\otimes I_{t_0},
\end{equation}
so Slater's condition holds, thus the dual supremum is always attained. 
\end{proof}

\subsection{Convex temporal-state optimal transport}
\label{subsec:convex-temporal-state-transport}

As proven in Ref.~\cite{jia2026unifiedspatiotemporalquantumstates}, we can take the probabilistic mixture of temporal state with difference branches to obtain a new temporal state, and this type of construction  unifies many existing frameworks of temporal states, including pseudo-density operator \cite{fitzsimons2015quantum} (or state over time \cite{fullwood2022quantum,Lie2025stateovertime}).

For the initial state $\rho_{t_0}$ and $n$ quantum channels with Choi matrices
${J}'_1,\cdots,{J}'_n$, let
$\operatorname{Branch}_n$ denote the set of allowed $2^n$ branch assignments.
The convex set of temporal states is defined as 
\begin{equation}
\begin{split}
\mathbf{State}_{T}
(\rho_{t_0};{J}'_1,\ldots,{J}'_n)
=
\operatorname{conv}
\Bigl\{
&
\Upsilon^{\mathtt{B}}
(\rho_{t_0};{J}'_1,\ldots,{J}'_n):
\mathtt{B}\in\operatorname{Branch}_n
\Bigr\}.
\end{split}
\label{eq:process-conditioned-branch-polytope}
\end{equation}
An element of this set has the form
\begin{equation}
\Upsilon
=
\sum_{\mathtt{B}\in\operatorname{Branch}_n}
p_{\mathtt{B}}\,
\Upsilon^{\mathtt{B}}
(\rho_{t_0};{J}'_1,\ldots,{J}'_n),
\label{eq:process-conditioned-branch-mixture}
\end{equation}
where
\begin{equation}
p_{\mathtt{B}}\geq0,
\quad
\sum_{\mathtt{B}}p_{\mathtt{B}}=1.
\end{equation}
In particular, the left,
right, mixed-branch, Margenau-Hill, and L\"uders-von Neumann temporal
states arise from distinguished choices of
$\{p_{\mathtt{B}}\}$ \cite{jia2026unifiedspatiotemporalquantumstates}. Since there are only $2^n$ branches for $t_0,\cdots,t_n$, $\mathbf{State}_{T}
(\rho_{t_0};{J}'_1,\ldots,{J}'_n)$ is indeed a convex polytope.

However, for temporal quantum optimal transport, we need to vary the dynamics, the above convex polytope of temporal state is thus not sufficient. 

\subsubsection{Fixed-branch convex temporal coupling set}

As we have see, fixed-branch temporal quantum optimal transport can be transformed into a multi-convex optimization problem over Choi operators $J'_k\in \mathbf{Choi}_k(\rho_{t_{k-1}}, \rho_{t_k})$. For two-time setting, this is equivalent to the  quantum channel optimal transport.
From temporal state perspective, we in fact consider the temporal coupling set as 
\begin{equation}
    \fC^{\ttB}(\rho_{t_0},\cdots,\rho_{t_{n}})=\{\Upsilon^{\mathtt{B}}
(\rho_{t_0};{J}'_1,\ldots,{J}'_n): J'_k\in \mathbf{Choi}_k(\rho_{t_{k-1},\rho_{t_k}}, k=1,\cdots,n\}.
\end{equation}
We can enlarge the coupling set by considering the convex hull
\begin{equation}
      \fC_{\rm conv}^{\ttB}(\rho_{t_0},\cdots,\rho_{t_{n}})=\operatorname{conv}  \fC^{\ttB}(\rho_{t_0},\cdots,\rho_{t_{n}}).
\end{equation}
Each state in $ \fC_{\rm conv}^{\ttB}(\rho_{t_0},\cdots,\rho_{t_{n}})$ gives correct marginals.
Take optimization over this set naturally give a convex temporal quantum optimal transport problem.
Also note that for two-time case, $  \fC_{\rm conv}^{\ttB}(\rho_{t_0},\cdots,\rho_{t_{1}})= \fC^{\ttB}(\rho_{t_0},\cdots,\rho_{t_{1}})$, since the set $ \mathbf{Choi}_1(\rho_{t_{0}}, \rho_{t_1})$ is convex and the initial state is fixed.

\subsubsection{General convex  temporal coupling set}

The more general convex temporal-state coupling set is the following:
\begin{equation}
\mathfrak{C}_{\mathrm{conv}}(\rho_{t_0},\cdots,\rho_{t_n})
=
\operatorname{conv}
\left(
\bigcup_{\mathtt{B}\in\operatorname{Branch}_n}
 \fC^{\ttB}(\rho_{t_0},\cdots,\rho_{t_{n}})
\right).
\label{eq:general-convex-temporal-state-set}
\end{equation}
Thus, temporal coupling in $\mathfrak{C}_{\mathrm{conv}}(\rho_{t_0},\cdots,\rho_{t_n})$ can be a probabilistic mixture of temporal states with different branches and with different Choi operators. Elements in $\mathfrak{C}_{\mathrm{conv}}(\rho_{t_0},\cdots,\rho_{t_n})$ gives correct marginals.
It's clear that $ \fC_{\rm conv}^{\ttB}(\rho_{t_0},\cdots,\rho_{t_{n}}) \subset \mathfrak{C}_{\mathrm{conv}}(\rho_{t_0},\cdots,\rho_{t_n})$, whic implies that
\begin{equation}
    \operatorname{QMK}_{\fC^{\ttB},\sC} \geq  \operatorname{QMK}_{\fC^{\ttB}_{\rm conv},\sC} \geq  \operatorname{QMK}_{\fC_{\rm conv},\sC}.
\end{equation}
The convex set of temporal Kirkwood-Dirac coupling is
\begin{equation}
\mathbb{KD}_{\mathrm{conv}}
(Q_{\rho_{t_0}},\cdots,Q_{\rho_{t_{n}}})
=
\left\{
Q_{\Upsilon}(\boldsymbol{\alpha})=\Tr[ F_{\boldsymbol{\alpha}} \Ups ]:
\Upsilon\in
\mathfrak{C}_{\mathrm{conv}}(\rho_{t_0},\cdots,\rho_{t_n})
\right\}.
\label{eq:convex-temporal-KD-plans}
\end{equation}
In temporal phase space, the problem can be formulated as
\begin{equation}
\begin{split}
\operatorname{QMK}_{\mathbb{KD}_{\mathrm{conv}}}
(Q_{\rho_{t_0}},\cdots,Q_{\rho_{t_{n}}})
&=
\inf_{Q\in
\mathbb{KD}_{\mathrm{conv}}
(Q_{\rho_{t_0}},\cdots,Q_{\rho_{t_{n}}})}
\left|
\sum_{\alpha_n,\ldots,\alpha_0}
C(\alpha_n,\ldots,\alpha_0)
Q(\alpha_n,\ldots,\alpha_0)
\right|
\nonumber\\
&=
\inf_{\Upsilon\in
\mathfrak{C}_{\mathrm{conv}}(\rho_{t_0},\cdots,\rho_{t_n})}
\left|
\Tr(\mathsf{C}\Upsilon)
\right|.
\end{split}
\label{eq:convex-temporal-MK}
\end{equation}

\subsubsection{Cost set in the complex plane and geometric duality}

The convex temporal quantum optimal transport can be transformed into a convex problem in complex plane.
Fix cost operator $\sC$, for each branch, define the set of attainable complex costs
\begin{equation}
\mathcal{C}_{\mathtt{B},\mathsf{C}}(\rho_{t_0},\cdots,\rho_{t_{n}})
=
\left\{
\Tr(\mathsf{C}\Upsilon):
\Upsilon\in
 \fC^{\ttB}(\rho_{t_0},\cdots,\rho_{t_{n}})
\right\}
\subset\mathbb{C},
\label{eq:branch-attainable-cost-set}
\end{equation}
and define
\begin{equation}
\mathcal{C}_{\mathrm{conv},\mathsf{C}}(\rho_{t_0},\cdots,\rho_{t_{n}})
=
\operatorname{conv}
\left(
\bigcup_{\mathtt{B}\in\operatorname{Branch}_n}
\mathcal{C}_{\mathtt{B},\mathsf{C}}
\right).
\label{eq:convex-attainable-cost-set}
\end{equation}

\begin{theorem}[Geometric duality]
\label{thm:temporal-cost-plane}
The convex temporal transport value is the Euclidean distance from the
origin to the convex attainable-cost set:
\begin{equation}
\operatorname{QMK}_{\mathrm{conv},\sC}
(\rho_{t_0},\cdots,\rho_{t_{n}})
=
\operatorname{dist}
\left(
0,\mathcal{C}_{\mathrm{conv},\mathsf{C}}(\rho_{t_0},\cdots,\rho_{t_{n}})
\right).
\label{eq:temporal-cost-plane-distance}
\end{equation}
Equivalently,
\begin{align}
\operatorname{QMK}_{\mathrm{conv},\sC}
(\rho_{t_0},\cdots,\rho_{t_{n}})
=
\sup_{|u|\leq1}
\inf_{\mathtt{B}\in\operatorname{Branch}_n}
\inf_{\Upsilon\in
\mathcal{C}_{\mathtt{B},\mathsf{C}}(\rho_{t_0},\cdots,\rho_{t_{n}})}
\operatorname{Re}
\left[
\overline{u}\Tr(\mathsf{C}\Upsilon)
\right].
\label{eq:convex-temporal-support-dual}
\end{align}
This provide a geometric duality for the original temporal optimal transport, and we can use it to simplify the optimization process.
\end{theorem}

\begin{proof}
By linearity of the cost functional, the set of attainable
costs over convex temporal couplings is precisely
$\mathcal{C}_{\mathrm{conv},\mathsf{C}}
(\rho_{t_0},\cdots,\rho_{t_n})$.
Consequently,
\begin{equation}
\begin{split}
&\operatorname{QMK}_{\mathrm{conv},\sC}
(\rho_{t_0},\cdots,\rho_{t_n})\\
&\quad=
\min_{z\in
\mathcal{C}_{\mathrm{conv},\mathsf{C}}
(\rho_{t_0},\cdots,\rho_{t_n})}
|z|\\
&\quad=
\operatorname{dist}
\left(
0,
\mathcal{C}_{\mathrm{conv},\mathsf{C}}
(\rho_{t_0},\cdots,\rho_{t_n})
\right),
\end{split}
\end{equation}
which proves \eqref{eq:temporal-cost-plane-distance}.

To establish the dual characterization, we use Eq.~\eqref{eq:modulus-support-function}.
Since
$\mathcal{C}_{\mathrm{conv},\mathsf{C}}
(\rho_{t_0},\cdots,\rho_{t_n})$
and the closed unit disk
$\{u\in\mathbb{C}:|u|\leq 1\}$
are nonempty compact convex sets, and
$\operatorname{Re}(\overline{u}z)$
is continuous and real-bilinear in $(u,z)$,
the minimax theorem yields
\begin{equation}
\begin{split}
&\inf_{z\in
\mathcal{C}_{\mathrm{conv},\mathsf{C}}
(\rho_{t_0},\cdots,\rho_{t_n})}
|z|\\
&\quad=
\sup_{|u|\leq 1}
\inf_{z\in
\mathcal{C}_{\mathrm{conv},\mathsf{C}}
(\rho_{t_0},\cdots,\rho_{t_n})}
\operatorname{Re}(\overline{u}z).
\end{split}
\end{equation}

For each fixed $u$, the functional
$z\mapsto\operatorname{Re}(\overline{u}z)$
is real-linear. Its infimum over a set therefore
coincides with its infimum over the convex hull
of that set. By the definition of the convex
attainable-cost set, we obtain
\begin{equation}
\begin{split}
&\inf_{z\in
\mathcal{C}_{\mathrm{conv},\mathsf{C}}
(\rho_{t_0},\cdots,\rho_{t_n})}
\operatorname{Re}(\overline{u}z)\\
&\quad=
\inf_{\mathtt{B}\in\operatorname{Branch}_n}
\inf_{\Upsilon\in
\mathcal{C}_{\mathtt{B},\mathsf{C}}
(\rho_{t_0},\cdots,\rho_{t_n})}
\operatorname{Re}
\left[
\overline{u}\Tr(\mathsf{C}\Upsilon)
\right].
\end{split}
\end{equation}
Combining these identities establishes
\eqref{eq:convex-temporal-support-dual}.
\end{proof}

\begin{corollary}[Criterion for a zero temporal transport cost]
\label{cor:zero-convex-temporal-cost}
Suppose that the convex attainable cost set is nonempty and compact.
Then
\begin{equation}
\operatorname{QMK}_{\mathrm{conv},\sC}
(\rho_{t_0},\cdots,\rho_{t_n})=0
\end{equation}
if and only if
\begin{equation}
0\in
\mathcal{C}_{\mathrm{conv},\mathsf{C}}
(\rho_{t_0},\cdots,\rho_{t_n}).
\label{eq:zero-cost-convex-hull-condition}
\end{equation}
Consequently, the convex temporal transport cost may vanish
even when the cost of every individual fixed branch is strictly
positive. This occurs when attainable complex costs from
different branches cancel under convex mixing.
\end{corollary}

\begin{proof}
By Theorem~\ref{thm:temporal-cost-plane},
\begin{equation}
\begin{split}
&\operatorname{QMK}_{\mathrm{conv},\sC}
(\rho_{t_0},\cdots,\rho_{t_n})\\
&\quad=
\operatorname{dist}
\left(
0,
\mathcal{C}_{\mathrm{conv},\mathsf{C}}
(\rho_{t_0},\cdots,\rho_{t_n})
\right).
\end{split}
\end{equation}
Since the convex attainable-cost set is compact and
therefore closed, its distance from the origin vanishes
if and only if it contains the origin.
\end{proof}

\begin{corollary}[Three temporal couplings suffice]
\label{cor:three-temporal-plans}
Suppose that the convex attainable cost set is nonempty
and compact, and that it is the convex hull of the
fixed-branch attainable costs.
For any fixed cost operator $\mathsf{C}$, the optimal
value of
$\operatorname{QMK}_{\mathrm{conv},\sC}
(\rho_{t_0},\cdots,\rho_{t_n})$
can be attained by a convex combination of at most
three fixed-branch temporal states.
If all attainable cost values are real, at most
two fixed-branch temporal states suffice.
\end{corollary}

\begin{proof}
Since
$\mathcal{C}_{\mathrm{conv},\mathsf{C}}
(\rho_{t_0},\cdots,\rho_{t_n})$
is nonempty and compact, there exists an optimal
cost value $z_*$ such that
\begin{equation}
\begin{split}
|z_*|
&=
\inf_{z\in
\mathcal{C}_{\mathrm{conv},\mathsf{C}}
(\rho_{t_0},\cdots,\rho_{t_n})}
|z|\\
&=
\operatorname{QMK}_{\mathrm{conv},\sC}
(\rho_{t_0},\cdots,\rho_{t_n}).
\end{split}
\end{equation}
By the definition of the convex attainable cost set,
$z_*$ belongs to the convex hull of the fixed-branch
attainable costs. Since these cost values lie in
$\mathbb{C}\cong\mathbb{R}^2$, Carath\'eodory's theorem
implies that there exist at most three fixed-branch
temporal states $\Upsilon_1,\Upsilon_2,\Upsilon_3$
and nonnegative coefficients $\lambda_1,\lambda_2,
\lambda_3$ satisfying
$\sum_{j=1}^{3}\lambda_j=1$, such that
\begin{equation}
z_*=
\sum_{j=1}^{3}
\lambda_j\Tr(\mathsf{C}\Upsilon_j).
\end{equation}
Thus, the convex combination
\begin{equation}
\Upsilon_*=
\sum_{j=1}^{3}\lambda_j\Upsilon_j
\end{equation}
attains the optimal convex temporal transport cost.
If all attainable cost values are real, the relevant
space has real dimension at most one, and
Carath\'eodory's theorem reduces the required
number of temporal states to two.
\end{proof}

The bound of three temporal states in
Corollary~\ref{cor:three-temporal-plans} concerns
the attainment of the optimal value for a fixed
complex cost operator. Representing an arbitrary
temporal state in the full operator space may
require more components.

\begin{corollary}[When convexification improves the optimum]
\label{cor:convexification-improvement}
For any fixed cost operator $\mathsf{C}$, the convex
temporal transport cost satisfies
\begin{equation}
\begin{split}
&\operatorname{QMK}_{\mathrm{conv},\sC}
(\rho_{t_0},\cdots,\rho_{t_n})\\
&\quad\leq
\inf_{\mathtt{B}\in\operatorname{Branch}_n}
\operatorname{QMK}_{\mathtt{B},\sC}
(\rho_{t_0},\cdots,\rho_{t_n}).
\end{split}
\label{eq:convex-temporal-improvement}
\end{equation}
The inequality can be strict due to cancellation
between complex cost values attained by different
branches.
By contrast, for the real-linear cost functional,
convexification does not change the optimal value:
\begin{equation}
\begin{split}
&\inf_{z\in
\mathcal{C}_{\mathrm{conv},\mathsf{C}}
(\rho_{t_0},\cdots,\rho_{t_n})}
\operatorname{Re}(z)\\
&\quad=
\inf_{\mathtt{B}\in\operatorname{Branch}_n}
\inf_{\Upsilon\in
\mathcal{C}_{\mathtt{B},\mathsf{C}}
(\rho_{t_0},\cdots,\rho_{t_n})}
\operatorname{Re}
\left[
\Tr(\mathsf{C}\Upsilon)
\right].
\end{split}
\label{eq:linear-cost-no-convexification-advantage}
\end{equation}
Thus, any improvement resulting from convexification
arises from the nonlinearity of the modulus cost.
\end{corollary}

\subsection{Dual branches, Hermitianization and Margenau-Hill temporal quantum optimal transport}

Let $\mathtt{B}^{c}$ be the complementary branch obtained by
interchanging $L$ and $R$ at every temporal slot. From the temporal
state construction,
\begin{equation}
\Upsilon^{\mathtt{B}^{c}}
=
\left(
\Upsilon^{\mathtt{B}}
\right)^\dagger.
\label{eq:dual-branch-adjoint}
\end{equation}

\begin{proposition}[Dual-branch symmetry]
\label{prop:dual-branch-transport-symmetry}
For every cost operator $\mathsf{C}$,
\begin{equation}
\operatorname{TMK}_{\mathtt{B},C}
(\boldsymbol{\rho})
=
\operatorname{TMK}_{\mathtt{B}^{c},C^\dagger}
(\boldsymbol{\rho}).
\label{eq:dual-branch-cost-symmetry}
\end{equation}
In particular, if $\mathsf{C}=\mathsf{C}^\dagger$, then
\begin{equation}
\operatorname{TMK}_{\mathtt{B},C}
(\boldsymbol{\rho})
=
\operatorname{TMK}_{\mathtt{B}^{c},C}
(\boldsymbol{\rho}).
\label{eq:Hermitian-dual-branch-symmetry}
\end{equation}
\end{proposition}

\begin{proof}
For corresponding temporal states on the two branches,
\begin{align}
\Tr\left[
\mathsf{C}^\dagger
\Upsilon^{\mathtt{B}^{c}}
\right]
&=
\Tr\left[
\mathsf{C}^\dagger
(\Upsilon^{\mathtt{B}})^\dagger
\right]
\nonumber\\
&=
\overline{
\Tr\left[
\mathsf{C}\Upsilon^{\mathtt{B}}
\right]
}.
\end{align}
Taking absolute values and then optimizing proves the result.
\end{proof}

A branch-symmetric mixture,
\begin{equation}
p_{\mathtt{B}}=p_{\mathtt{B}^{c}},
\label{eq:branch-symmetric-probability}
\end{equation}
produces a Hermitian temporal state, called Margenau-Hill state \cite{Jia2025TemporalKirkwoodDirac,jia2026unifiedspatiotemporalquantumstates}. For example
\begin{equation}
\frac{1}{2}
\left(
\Upsilon^{\mathtt{B}}
+
\Upsilon^{\mathtt{B}^{c}}
\right)
=
\frac{1}{2}
\left(
\Upsilon^{\mathtt{B}}
+
(\Upsilon^{\mathtt{B}})^\dagger
\right)
\label{eq:branch-Hermitianization}
\end{equation}
is a Margenau-Hill temporal state. For a Hermitian
cost, its cost value is
\begin{equation}
\Tr\left[
\mathsf{C}
\frac{
\Upsilon^{\mathtt{B}}
+
\Upsilon^{\mathtt{B}^{c}}
}{2}
\right]
=
\operatorname{Re}
\Tr\left[
\mathsf{C}\Upsilon^{\mathtt{B}}
\right].
\label{eq:MH-real-cost}
\end{equation}
The L\"uders-von Neumann state (pseudo-density operator) is obtained by a distinguished
branch-symmetric mixture over all admissible orderings,
\begin{equation}
    \Ups=\frac{1}{2^n}\sum_{\ttB\in \operatorname{Branch}_n} \Ups^{\ttB}.
\end{equation}
It has a recursive expression
\begin{equation}
    \Ups_{t_k,\cdots,t_0}=\frac{1}{2}\{J[\cE_{k}], \Ups_{t_{k-1},\cdots,t_0}\}, \Ups_{t_0}=\rho_{t_0},
\end{equation}
where $\{\bullet,\bullet\}$ is anti-commutator.
The optimal transport for pseudo-density operator are recently discussed in \cite{hoogstederriera2026PDOMK}, which can be regarded as a special case of our setting.
From our previous discussion and combining the Hermitization, we will obtain more temporal optimal transport with Hermitian temporal states.

\begin{figure}
    \centering
    \includegraphics[width=0.5\linewidth]{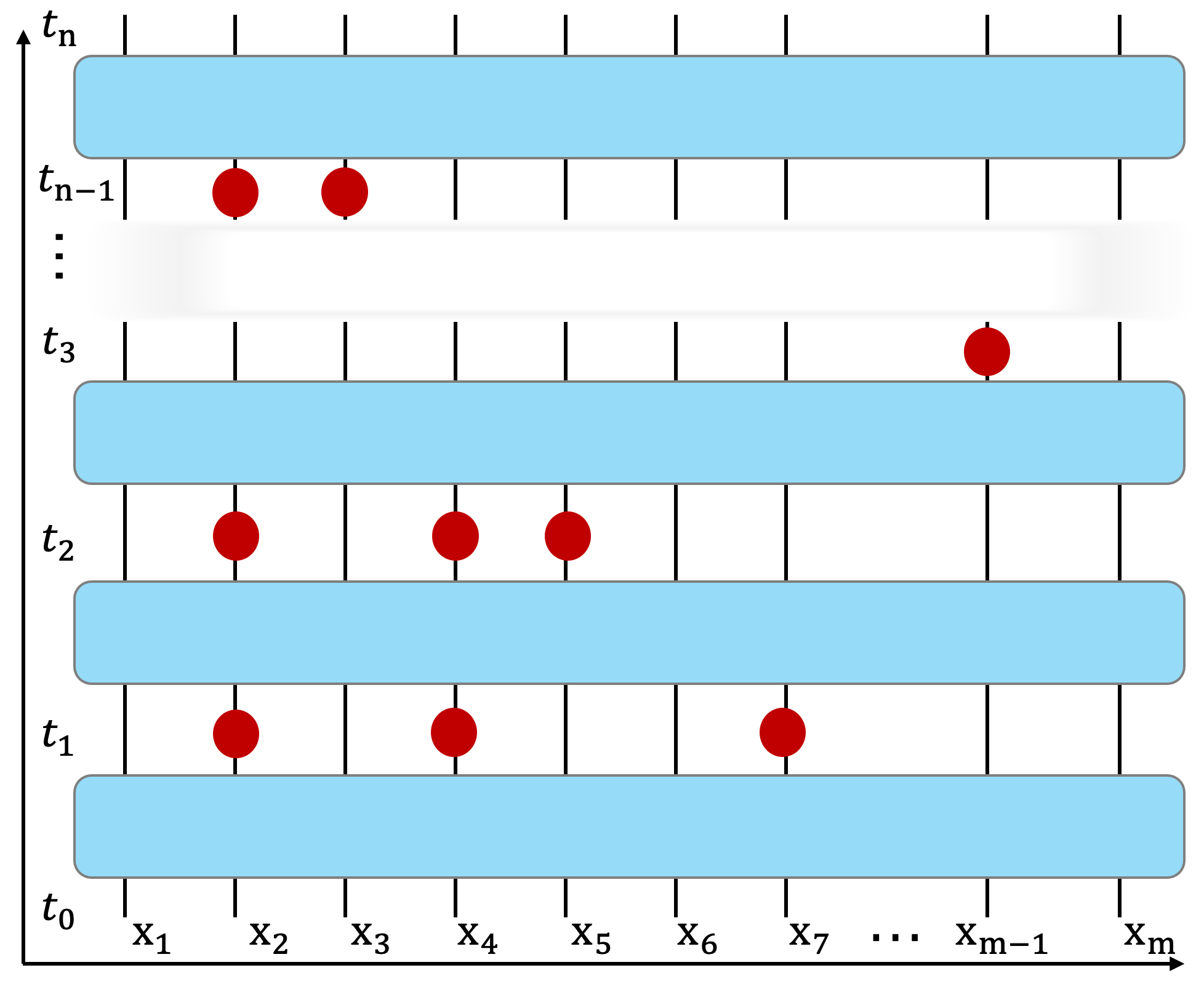}
    \caption{The background spacetime lattice and spacetime set.}
    \label{fig:STlattice}
\end{figure}

\section{Spatiotemporal optimal transport with a fixed background structure}
\label{sec:SToptoutline}
The general spatiotemporal optimal transport requires further structure. Since, given an arbitrary spacetime set $\cA$, the set of all spatiotemporal states is not well defined, we need to introduce a background structure \cite{jia2023quantumspace}.

We will consider the following setting, given a background spacetime lattice 
$$\mathcal{L}=\{(t_0,x_1),\cdots, (t_0,x_m);\cdots ; (t_n,x_1),\cdots, (t_n,x_m)\}.$$
A spacetime set $\cA\subset \cL$, see Figure~\ref{fig:STlattice} for illustration, where the spacetime set is shown as red dots.
For the spacetime lattice $\cL$, the spatiotemporal state is well-defined, which are determined by initial state $\rho_{x_1,\cdots,x_m}(t_0)$ and evolutions $\mathcal{E}_{t_k\rightarrow t_{k+1}}$, $k=0,\cdots,n-1$. For instance, for fixed branch $\ttB$, it is of the form 
\begin{equation}
    \Ups_{\cL}^{\ttB}=J_n \star_{\ttB} J_{n-1}  \star_{\ttB} \cdots  \star_{\ttB} J_1  \star_{\ttB} \rho_{x_1,\cdots,x_m}(t_0),
\end{equation}
where $J_k$'s are Jami{\l}kowski operators. The spatiotemporal state for $\cA$ is thus obtained by taking reduced state $\Ups_{\cA}=\Tr_{\cL\setminus \cA}\Ups_{\cL}$. Since $\Ups_{\cL}$ is a temporal state, the discussion in the previous section can be applied, we can first consider the temporal couplings for $\cL$ and then take the reduced state to define the quantum optimal transport for the general spatiotemporal setting. 
The optimization will taken over the initial state $\rho_{x_1,\cdots,x_m}(t_0)$ and all quantum channels $\mathcal{E}_{t_k\rightarrow t_{k+1}}$, $k=0,\cdots,n-1$ that gives the correct marginals.

Take fixed branch case as an example, the set of all spatiotemporal state for $\cL$ is
\begin{equation}
\begin{split}
  \mathbf{State}^{\ttB}(\cL)= \{&\Ups=J_n \star_{\ttB} J_{n-1}  \star_{\ttB} \cdots  \star_{\ttB} J_1  \star_{\ttB} \rho_{x_1,\cdots,x_m}(t_0):\nonumber\\
&
J_k \in \mathbf{Jami}, \rho_{x_1,\cdots,x_m}(t_0) \geq 0, \Tr \rho_{x_1,\cdots,x_m}(t_0)=1\}
\end{split}
\label{eq:fixed-branch-temporal-state-setL}
\end{equation}
The set of all spatiotemporal state for $\cA$ is $\mathbf{State}^{\ttB}(\cA)=\{\Ups_{\cA}=\Tr_{\cL\setminus \cA}\Ups_{\cL}:\Ups\in \mathbf{State}^{\ttB}(\cL)\}$
For a set of given marginal states $\{\rho_{z_k}\}$, the spatiotemporal couplings are thus
\begin{equation}
    \mathfrak{C}^{\ttB}(\rho_{z_1},\cdots,\rho_{z_N}) =\{ \Ups\in \mathbf{State}^{\ttB}(\cA): \Tr_{\widehat{z_k}} \Ups =\rho_{z_k}\}
\end{equation}
All other types of couplings in the previous section can be defined similarly.

\section{Kantorovich-Rubinstein metric for spatiotemporal states}
\label{sec:KRmetric}
The phase-space description makes it more natural to define the quantum Wasserstein distance, since it allows for a direct comparison with the classical definition. See Ref.~\cite[Chapter 6]{villani2009optimal} for basics of Wasserstein distance. Recall that for classical probability distributions $p$ and $q$ over the same polish space $\mathcal{X}$, the Kantorovich-Rubinstein theorem states that the Wasserstein-1 distance can be expressed in the Kantorovich-Rubinstein form using $1$-Lipschitz functions: $W_1(p,q)=\operatorname{KR}(p,q)$.  In this section, we generalize this to spatiotemporal states. We emphasize that our approach is quite different from those discussed in spatial quantum optimal transport, e.g., \cite{Toth2023quantumwasserstein,De2023Wasserstein,chakrabarti2019quantum,DePalma2021Wasserstein}, where they use a generalization of Wasserstein's definition.

For spatiotemporal causal set $\cA=\{z_1,\cdots,z_N\}$ with $z_k=(t_k,x_k)$'s spacetime points, let
$\Upsilon,\Xi\in\mathbf{State}_{\rm st}(\mathcal{H}_{\cA})$ be two
spatiotemporal states on the same lattice. Their IC-POVM phase-space
representations are
\begin{equation}
    Q_\Upsilon(\boldsymbol{\alpha})
    =
    \Tr[F_{\boldsymbol{\alpha}}\Upsilon],
    \quad
    Q_\Xi(\boldsymbol{\alpha})
    =
    \Tr[F_{\boldsymbol{\alpha}}\Xi].
\end{equation}
Both are normalized:
\begin{equation}
    \sum_{\boldsymbol{\alpha}}
    Q_\Upsilon(\boldsymbol{\alpha})
    =
    \sum_{\boldsymbol{\alpha}}
    Q_\Xi(\boldsymbol{\alpha})
    =
    1.
\end{equation}
Their difference therefore has total mass zero, although it can be
complex.

The phase-space is $\Omega_{\cA}=\Omega_{z_1}\times \cdots \times \Omega_{z_N}$, we can introduce equip a \emph{ground distance} $d_{\cA}(\boldsymbol{\alpha},\boldsymbol{\beta})$ on it.
For a function
$f:\Omega_{\cA}\rightarrow\mathbb{C}$, define Lipschitz seminorm
\begin{equation}
    \operatorname{Lip}_{d_{\cA}}(f)
    =\|f\|_{\rm Lip}=
    \sup_{\boldsymbol{\alpha}\neq\boldsymbol{\beta}}
    \frac{
    |f(\boldsymbol{\alpha})-f(\boldsymbol{\beta})|
    }{
    d_{\cA}(\boldsymbol{\alpha},\boldsymbol{\beta})
    }.
\end{equation}
Equivalently, \(\operatorname{Lip}(f)\) is the smallest number \(L \ge 0\) such that
\[
|f(x)-f(y)| \le L\, d(x,y)
\quad \text{for all } x,y \in X.
\]
If \(\operatorname{Lip}(f) \le 1\), then \(f\) is called \(1\)-Lipschitz.
It is called a seminorm since $\|f\|_{\rm Lip}=0$ does not imply $f=0$, but all other conditions of norm are satisfied.

\begin{definition}[Spatiotemporal Kantorovich–Rubinstein distance]

For two spatiotemporal state $\Upsilon,\Xi\in \mathbf{State}(\cH_{\cA})$, and a chosen IC-POVM spatiotemporal frame $\{F_{\boldsymbol{\alpha}}\}$, fix a reference point
$\boldsymbol{\alpha}_0\in\Omega_{\cA}$. The spatiotemporal
Kantorovich–Rubinstein distance induced by spatiotemporal Kirkwood-Dirac distribution is
\begin{align}
\operatorname{KR}(\Upsilon,\Xi)
=
\sup
\Biggl\{
&
\left|
\sum_{\boldsymbol{\alpha}\in \Omega_{\cA}}
f(\boldsymbol{\alpha})
\left[
Q_\Upsilon(\boldsymbol{\alpha})
-
Q_\Xi(\boldsymbol{\alpha})
\right]
\right|:
\nonumber\\
&
f:\Omega_{\cA}\to \Cbb,\,\operatorname{Lip}_{d_{\cA}}(f)\leq1,
\,
f(\boldsymbol{\alpha}_0)=0
\Biggr\}.
\label{eq:ST-KR-metric}
\end{align}
The condition $f(\boldsymbol{\alpha}_0)=0$ removes the irrelevant
freedom to add a constant, as given constant $c$, $f_c(\bullet):=f+c$ has the same Lipschitz norm as $f$, $f(\boldsymbol{\alpha}_0)=0$ chooses exactly one representative from each equivalence class $\{f+c\}$.
For the Margenau-Hill distribution, we may restrict $f$ to real-valued functions, thereby obtaining $\operatorname{KR}_{\rm MH}(\Upsilon,\Xi)$.
We can also define the Kantorovich-Rubinstein distance in Eq.~\eqref{eq:ST-KR-metric} by taking the supremum over all real-valued functions. To distinguish the two cases, we denote them by $\operatorname{KR}^{\Rbb}_{\rm KD}(\Upsilon,\Xi)$ and $\operatorname{KR}_{\rm KD}^{\Cbb}(\Upsilon,\Xi)$. Note that $\operatorname{KR}^{\Rbb}_{\rm KD}(\Upsilon,\Xi) \ge \operatorname{KR}_{\rm MH}(\Upsilon,\Xi)$.
\end{definition}

The metric also has the operator representation
\begin{align}
\operatorname{KR}(\Upsilon,\Xi)
=
\sup_{\substack{
\operatorname{Lip}_{d_{\cA}}(f)\leq1\\
f(\boldsymbol{\alpha}_0)=0}}
\left|
\Tr
\left[
A_f(\Upsilon-\Xi)
\right]
\right|,
\label{eq:ST-KR-operator-form}
\end{align}
where
\begin{equation}
    A_f
    =
    \sum_{\boldsymbol{\alpha}}
    f(\boldsymbol{\alpha})F_{\boldsymbol{\alpha}}.
\end{equation}
For $f\in \Rbb^{\Omega_{\cA}}$, $A_f$ is Hermitian.
The Kantorovich–Rubinstein distance measures the largest distinguishability of the two
spatiotemporal states by a phase-space observable whose coefficients is
1-Lipschitz.

\begin{theorem}[Metric properties]\label{thm:KRmetric}
For a finite metric phase space and a tensor product of local
IC-POVM frames, $\operatorname{KR}(\Upsilon,\Xi)$ is a metric on
$\mathbf{State}(\cH_{\cA})$ and a metric on $\mathbf{KD}(\cA)$ if regard as function of $Q_{\Upsilon}$ and $Q_{\Xi}$.
\end{theorem}

\begin{proof}
Nonnegativity is immediate. Exchanging $\Upsilon$ and $\Xi$ changes
the sign of the expression inside the absolute value, so symmetry
holds.

For three spatiotemporal states $\Upsilon,\Xi,\Theta$, every admissible
$f$ satisfies
\begin{align}
&
\left|
\sum_{\boldsymbol{\alpha}}
f(\boldsymbol{\alpha})
\left[
Q_\Upsilon(\boldsymbol{\alpha})
-
Q_\Theta(\boldsymbol{\alpha})
\right]
\right|
\nonumber\\
&\quad\leq
\left|
\sum_{\boldsymbol{\alpha}}
f(\boldsymbol{\alpha})
\left[
Q_\Upsilon(\boldsymbol{\alpha})
-
Q_\Xi(\boldsymbol{\alpha})
\right]
\right|
+
\left|
\sum_{\boldsymbol{\alpha}}
f(\boldsymbol{\alpha})
\left[
Q_\Xi(\boldsymbol{\alpha})
-
Q_\Theta(\boldsymbol{\alpha})
\right]
\right|.
\end{align}
Taking the supremum gives the triangle inequality.
It remains to prove definiteness. If
\begin{equation}
   \operatorname{KR}(\Upsilon,\Xi)=0,
\end{equation}
then the difference $Q_\Upsilon-Q_\Xi$ pairs to zero with every
Lipschitz function. On a finite metric space (like $\Omega_{\cA}$) every function is
Lipschitz after multiplication by a sufficiently small nonzero
constant. Hence
\begin{equation}
    Q_\Upsilon(\boldsymbol{\alpha})
    =
    Q_\Xi(\boldsymbol{\alpha})
\end{equation}
for every $\boldsymbol{\alpha}$. Informational completeness of the
tensor-product frame implies $\Upsilon=\Xi$.
\end{proof}

When $Q_\Upsilon$ and $Q_\Xi$ are positive probability distributions,
the complex dual formulation
\eqref{eq:ST-KR-metric} reduces to the classical
Kantorovich-Rubinstein metric which coincides with classical $1$-Wasserstein metric.

For spatial density operators $\rho_A,\rho_B \in \State_{\rm s}(\cH)$, and IC-POVM frame $\{F_{\boldsymbol{\alpha}}, F_{\boldsymbol{\beta}}\}$
$Q_{\rho_A}(\boldsymbol{\alpha})$ and $Q_{\rho_B}(\boldsymbol{\beta})$ are probability distributions,  give a ground distance $d(\bfalpha,\bfbeta)$ over phase space and take supremum over real 1-Lipschitz functions, we obtain a distance function of density operators $\KR(\rho_A,\rho_B):=\KR^{\Rbb}_{\rm KD}(\rho_A,\rho_B)=\KR_{\rm MH}(\rho_A,\rho_B)$. 
Recall that classical $p$-Wasserstein metric is defined by
\begin{equation}
    W^C_p(\rho_A,\rho_B)=\left[\inf\Biggl\{ \sum_{\bfalpha,\bfbeta} d(\bfalpha,\bfbeta)^p P(\bfalpha,\bfbeta):\sum_{\bfbeta}P(\bfalpha,\bfbeta)= Q_{\rho_A}(\boldsymbol{\alpha}),\sum_{\bfalpha}P(\bfalpha,\bfbeta)= Q_{\rho_B}(\boldsymbol{\beta}) \Biggr\}\right]^{1/p}
\end{equation}
We have $\KR(\rho_A,\rho_B)=W_1^C(\rho_A,\rho_B)$, Theorem~\ref{thm:KRmetric} implies that this is a metric.

Assume that the joint frame is the product frame
$\{F_{\boldsymbol{\alpha}}\otimes F_{\boldsymbol{\beta}}\}$, so that
\begin{equation}
Q_{\rho_{AB}}(\boldsymbol{\alpha},\boldsymbol{\beta})
=
\Tr[
(F_{\boldsymbol{\alpha}}\otimes F_{\boldsymbol{\beta}})
\rho_{AB}
].
\end{equation}
Every such distribution is a classical coupling of
$Q_{\rho_A}$ and $Q_{\rho_B}$. However, not every classical coupling
admits this quantum realization.
We can also define
\begin{equation}
    W^Q_p(\rho_A,\rho_B)=\left[\inf\Biggl\{ \sum_{\bfalpha,\bfbeta} d(\bfalpha,\bfbeta)^p Q_{\rho_{AB}}(\bfalpha,\bfbeta):\Tr_B  \rho_{AB}=\rho_A,\Tr_A  \rho_{AB}=\rho_B\Biggr\}\right]^{1/p}.
\end{equation}
It is generally not a metric.
To see this, consider the qubit tetrahedral informationally
complete positive operator-valued measure
\begin{equation}
F_i=\frac14(\mathds{1}+\boldsymbol{v}_i\cdot\boldsymbol{\sigma}),
\quad i=1,\ldots,4,
\end{equation}
and the discrete metric $d(i,j)=1-\delta_{ij}$.
For every $p>0$, we have $d(i,j)^p=d(i,j)$. Consequently,
\begin{align}
W_p^Q(\mathds{1}/2,\mathds{1}/2)
&=
\left[1-
\sup_{\substack{\rho_{AB}\geq0\\
\Tr_B\rho_{AB}=\Tr_A\rho_{AB}=\mathds{1}/2}}
\Tr\left[\rho_{AB}\sum_iF_i\otimes F_i\right] \right]^{1/p}
\nonumber\\
&=
(1-\frac13)^{1/p}
=(\frac23)^{1/p},
\end{align}
where we used\footnote{
Recall that tetrahedral unit vectors are
\begin{equation}
\boldsymbol{v}_1=\frac{(1,1,1)}{\sqrt3},\quad
\boldsymbol{v}_2=\frac{(1,-1,-1)}{\sqrt3},\quad
\boldsymbol{v}_3=\frac{(-1,1,-1)}{\sqrt3},\quad
\boldsymbol{v}_4=\frac{(-1,-1,1)}{\sqrt3}.
\end{equation}
they satisfy
\begin{equation}
\sum_i\boldsymbol{v}_i=0,
\quad
\sum_i\boldsymbol{v}_i\boldsymbol{v}_i^{\mathsf T}
=\frac43 \mathds{1}_3.
\end{equation}
Each coordinate sums to zero, each diagonal second
moment equals $4/3$, and each off-diagonal second moment vanishes.
Expanding the tensor products therefore gives
\begin{align}
\sum_{i=1}^4F_i\otimes F_i
&=\frac1{16}\sum_{i=1}^4
\Bigl[
\mathds{1}\otimes \mathds{1}
+(\boldsymbol{v}_i\cdot\boldsymbol{\sigma})\otimes \mathds{1}
+\mathds{1}\otimes(\boldsymbol{v}_i\cdot\boldsymbol{\sigma})
\nonumber\\
&\hspace{35mm}
+\sum_{a,b=1}^3v_{i,a}v_{i,b}\,
\sigma_a\otimes\sigma_b
\Bigr]
\nonumber\\
&=\frac14\mathds{1}\otimes \mathds{1}
+\frac1{12}\sum_{a=1}^3\sigma_a\otimes\sigma_a.
\end{align}
For two qubits, the swap operator satisfies
\begin{equation}
\mathbb{S}=\frac12\left(
\mathds{1}\otimes \mathds{1}+\sum_{a=1}^3\sigma_a\otimes\sigma_a
\right).
\end{equation}
Hence
\begin{align}
\sum_{i=1}^4F_i\otimes F_i
&=\frac14\mathds{1}\otimes \mathds{1}+\frac1{12}(2\mathbb{S}-\mathds{1}\otimes \mathds{1})
\nonumber\\
&=\frac16(\mathds{1}\otimes \mathds{1}+\mathbb{S})
=\frac13P_{\mathrm{sym}},
\end{align}
where $P_{\mathrm{sym}}=(\mathds{1}\otimes \mathds{1}+\mathbb{S})/2$.}
\begin{equation}
\sum_iF_i\otimes F_i=\frac13P_{\mathrm{sym}},
\end{equation}
with $P_{\mathrm{sym}}=(\mathds{1}\otimes \mathds{1} +\mathbb{S})/2$ orthogonal projector onto the symmetric subspace,  where $\mathbb{S}$ is swap operator. 
The supremum is attained by
$\rho_{AB}=P_{\mathrm{sym}}/3$, whose two marginals are $\mathds{1}/2$.
Thus $W_p^Q(\rho,\rho)$ need not vanish, violating a necessary
metric axiom.

\begin{proposition}
For a finite metric outcome space and a product informationally
complete positive operator-valued measure frame,
\begin{equation}
\mathrm{KR}(\rho_A,\rho_B)=W^C_1(\rho_A,\rho_B) \leq W^Q_1(\rho_A,\rho_B).
\end{equation}
Equality holds if and only if at least one optimal classical
transport plan is induced by a bipartite density operator with
marginals $\rho_A$ and $\rho_B$.
\end{proposition}

\begin{proof}
Let $\rho_{AB}$ be any admissible bipartite density operator and
write $\pi(\boldsymbol{\alpha},\boldsymbol{\beta})
=Q_{\rho_{AB}}(\boldsymbol{\alpha},\boldsymbol{\beta})$.
For every real-valued $1$-Lipschitz function $f$,
\begin{align}
&\left|
\sum_{\boldsymbol{\alpha}}
f(\boldsymbol{\alpha})
[Q_{\rho_A}(\boldsymbol{\alpha})
-Q_{\rho_B}(\boldsymbol{\alpha})]
\right|
\nonumber\\
&\quad=
\left|
\sum_{\boldsymbol{\alpha},\boldsymbol{\beta}}
[f(\boldsymbol{\alpha})-f(\boldsymbol{\beta})]
\pi(\boldsymbol{\alpha},\boldsymbol{\beta})
\right|
\nonumber\\
&\quad\leq
\sum_{\boldsymbol{\alpha},\boldsymbol{\beta}}
d(\boldsymbol{\alpha},\boldsymbol{\beta})
\pi(\boldsymbol{\alpha},\boldsymbol{\beta}).
\end{align}
Taking the supremum over $f$ and then the infimum over
$\rho_{AB}$ proves the inequality.

Classical Kantorovich--Rubinstein duality identifies
$\mathrm{KR}(\rho_A,\rho_B)$ with the minimum cost over all
classical couplings of $Q_{\rho_A}$ and $Q_{\rho_B}$.
The admissible quantum states form a compact set, and their
induced distributions form a subset of these classical couplings.
Thus equality holds precisely when this subset contains a
classical optimal coupling.
\end{proof}

For example, consider a qubit with the tetrahedral
informationally complete positive operator-valued measure, we have
\begin{equation}
\mathrm{KR}(\mathds{1}/2,\mathds{1}/2)=0.
\end{equation}
But
\begin{equation}
W_1^Q(\mathds{1}/2,\mathds{1}/2)=\frac23
>
\mathrm{KR}(\mathds{1}/2,\mathds{1}/2)=0.
\end{equation}

\section{Conclusions and outlook}

In this work, we establish a framework of quantum optimal transport in spacetime using the Kirkwood-Dirac phase space and spatiotemporal state. We carefully discussed the duality and Kantorovich-Rubinstein metric.

Several problems are worth further investigation. First, a general definition of the spatiotemporal Wasserstein $p$-metric for $p>1$ has not yet been established and thus requires further study.  
Second, the continuous-variable case remains largely unexplored: the corresponding phase-space formalism can be developed by directly extending our finite-dimensional discussion to infinite dimensions, but a spatiotemporal-state formalism requires further work. This is because that the Choi and Jamio{\l}kowski operators are not well-defined in this case, so a different approach needs to be developed. 
Third, one can also consider subsets of couplings that give rise to different correlations, which would yield a hierarchy of spatiotemporal optimal transport. For instance, consider the spatial correlation hierarchy: quantum discord, quantum entanglement, quantum steering, and Bell nonlocality (see \cite{Toth2023quantumwasserstein} for application in entanglement theory and \cite{Shi2024Coherence} for applications in quantum coherence). As the coupling set shrinks, the optimal transport value increases; extending this to general spatiotemporal correlations may shed new light on the study of different types of correlation in the case where only local data are available.
Finally, the connection of our formalism with those discussed in gauge/gravity duality \cite{hashimoto2026holographyoptimaltransportemergent} and in quantum chaos \cite{hashimoto2026wassersteinspacequantumchaos} is also a crucial open direction.
These will be addressed in our future work.

\begin{acknowledgments}
Z.J. is supported by by the National Natural Science Foundation of China (Grant
No. 12605015), Start-up Grant of Central South University (Grant No. 502045031) and by the Frontier Interdisciplinary Direction ``Quantum Technologies'' Project of Central South University (Grant No. 506010805). 
\end{acknowledgments}

\appendix

\section{Some technical results from convex analysis and matrix analysis}

In this appendix, we recall some technical results from convex analysis and matrix analysis that will be useful for the discussion in the main text.

\begin{theorem}[Carath\'eodory theorem~\cite{Caratheodory1911}]
Let $A \subseteq \mathbb{R}^{d}$. Then every point in the
convex hull $\operatorname{conv}(A)$ can be written as a
convex combination of at most $d+1$ points of $A$.
\end{theorem}

\begin{theorem}[Slater's theorem for semidefinite programming~\cite{watrous2012simpler}]
\label{thm:SDPslater}
The following implications hold for every semidefinite programming \((\mathcal{E}, C, D)\). Let 
\[
\mathcal{P} = \{X \in \mathbf{Pos}(\mathcal{X}) : \mathcal{E}(X) = D\} \quad \text{and} \quad \mathcal{D} = \{Y \in \mathbf{Herm}(\mathcal{Y}) : \mathcal{E}^{\dagger}(Y) \geq C\}
\]
denote the sets of primal and dual feasible operators, and $\alpha$ and $\beta$ denote the primal and dual optimal values respectively.
\begin{enumerate}
    \item If \(\mathcal{P} \neq \varnothing\) and there exists a Hermitian operator \(Y\) for which \(\mathcal{E}^{\dagger}(Y) > C\), then the primal and dual optimal values coincide and there exists a primal feasible operator \(X \in \mathcal{P}\) for which \(\langle C, X \rangle = \alpha\).
    \item If \(\mathcal{D} \neq \varnothing\) and there exists a positive definite operator \(X > 0\) for which \(\mathcal{E}(X) = D\), then the primal and dual optimal values coincide and there exists a dual feasible operator \(Y \in \mathcal{D}\) for which \(\langle D, Y \rangle = \beta\).
\end{enumerate}
\end{theorem}

\begin{theorem}[Sion's minimax theorem~\cite{sion1958general}]\label{thm:minimax}
Let \(\mathcal{X}\) be a compact convex subset of a topological vector space, and let \(\mathcal{Y}\) be a convex subset of a topological vector space. If \(f: \mathcal{X} \times \mathcal{Y} \to \mathbb{R}\) is quasi-concave and upper semicontinuous in \(x\) for each fixed \(y \in \mathcal{Y}\), and quasi-convex and lower semicontinuous in \(y\) for each fixed \(x \in \mathcal{X}\), then
\[
\sup_{x \in \mathcal{X}} \inf_{y \in \mathcal{Y}} f(x, y) = \inf_{y \in \mathcal{Y}} \sup_{x \in \mathcal{X}} f(x, y).
\]
\end{theorem}

\begin{proposition}[Cyclicity of the partial transpose]\label{lemma:PartialTranspose}
Let $\mathcal{H}=\mathcal{H}_A\otimes\mathcal{H}_B$ and let $X,Y\in\mathbf{B}(\mathcal{H})$. 
Denote by $T_B$ the partial transpose on subsystem $B$. Then
\begin{equation}
\operatorname{Tr}\!\left(X^{T_B}Y\right)=\operatorname{Tr}\!\left(X\,Y^{T_B}\right).
\label{eq:partial-transpose-cyclic}
\end{equation}
\end{proposition}

\begin{proof}
Choose product bases $\{|i\rangle_A\}$ and $|\alpha\rangle_B$ and write
$X=\sum_{i,j,\alpha,\beta}X_{i\alpha,j\beta}\,|i\rangle\langle j|\otimes|\alpha\rangle\langle\beta|$.
By definition of the partial transpose,
$\big(X^{T_B}\big)_{i\alpha,j\beta}=X_{i\beta,j\alpha}$.
Hence
\begin{align*}
\operatorname{Tr}\!\left(X^{T_B}Y\right)
&=\sum_{i,\alpha,j,\beta}\big(X^{T_B}\big)_{i\alpha,j\beta}\,Y_{j\beta,i\alpha}
 =\sum_{i,\alpha,j,\beta}X_{i\beta,j\alpha}\,Y_{j\beta,i\alpha},\\
\operatorname{Tr}\!\left(X\,Y^{T_B}\right)
&=\sum_{i,\alpha,j,\beta}X_{i\alpha,j\beta}\,\big(Y^{T_B}\big)_{j\beta,i\alpha}
 =\sum_{i,\alpha,j,\beta}X_{i\alpha,j\beta}\,Y_{j\alpha,i\beta}.
\end{align*}
The two expressions coincide after relabelling the dummy indices $\alpha\leftrightarrow\beta$ in the first line. 
\end{proof}

\begin{proposition}
    For operator $X=\sum_{ij}X_{ij}|i\rangle\langle j|$, let $|X\rrangle=\sum_{ij}X_{ij}|i\rangle|j\rangle$ and $|\Omega\rangle=\sum_i |ii\rangle$, we have 
    \begin{equation}
    \begin{aligned}
         |X\rrangle = (X\otimes \mathds{1}) |\Omega\rangle = (\mathds{1}\otimes X^T)|\Omega\rangle,\\
         \llangle X| = \llangle\Omega|(X^{\dagger} \otimes \mathds{1}) = \llangle\Omega|(\mathds{1} \otimes X^{*}).
    \end{aligned}\label{eq:vectorization}
    \end{equation}
\end{proposition}

\bibliographystyle{apsrev4-1-title}
\bibliography{Jiabib}
\end{document}

%% file: MathPhysDef.tex
\usepackage{MathPhystitle}
\pdfoutput = 1
\usepackage[T1]{fontenc}

\usepackage{amsmath}
\usepackage{amssymb}
\usepackage{amsthm}
\usepackage{bm}
\usepackage{braket}
\usepackage{graphicx}
\usepackage{adjustbox}

\usepackage{mathdots}

\usepackage{multirow}
\usepackage{tabu}

\usepackage{booktabs} 
\usepackage{xcolor}   
\usepackage{hhline}   

\usepackage{orcidlink} 

\usepackage{tikz}
\usetikzlibrary{positioning,intersections}
\usetikzlibrary{calc}
\usetikzlibrary{shapes.geometric}
\usepackage{braids}
\usepackage{tikz-cd}
\usetikzlibrary{shapes.geometric,shapes.misc}
\usetikzlibrary{arrows,matrix,calc,scopes,decorations.markings,snakes}
\usetikzlibrary{arrows.meta}
\usetikzlibrary{intersections, patterns,fit} 
\usetikzlibrary{decorations.pathreplacing,angles,quotes}
\usetikzlibrary{tqft} 

\definecolor{darkblue}{RGB}{40, 85, 120} 

\newtheorem{theorem}{Theorem}[section]

\newtheorem{proposition}[theorem]{Proposition}
\newtheorem{corollary}[theorem]{Corollary}

\theoremstyle{definition}
\newtheorem{definition}[theorem]{Definition}

\theoremstyle{remark}

\usepackage[bbgreekl]{mathbbol} 

\usepackage{dsfont}

\usepackage{microtype}       

\DeclareMathAlphabet{\mathcal}{OMS}{cmsy}{m}{n}

\usepackage{euscript}

\newcommand{\Rbb}{\mathbb{R}}

\newcommand{\Cbb}{\mathbb{C}}

\newcommand{\Herm}{\mathbf{Herm}}
\newcommand{\State}{\mathbf{State}}

\newcommand{\bfalpha}{\boldsymbol{\alpha}}
\newcommand{\bfbeta}{\boldsymbol{\beta}}

\newcommand{\KR}{\operatorname{KR}}

\newcommand\Tr{\operatorname{Tr}}

\newcommand{\id}{\operatorname{id}}

\newcommand{\cA}{\text{\usefont{OMS}{cmsy}{m}{n}A}}

\newcommand{\cE}{\text{\usefont{OMS}{cmsy}{m}{n}E}}

\newcommand{\cH}{\text{\usefont{OMS}{cmsy}{m}{n}H}}

\newcommand{\cL}{\text{\usefont{OMS}{cmsy}{m}{n}L}}

\newcommand{\cX}{\text{\usefont{OMS}{cmsy}{m}{n}X}}
\newcommand{\cY}{\text{\usefont{OMS}{cmsy}{m}{n}Y}}

\newcommand\sC           {\mathsf{C}}

\newcommand\fC           {\mathfrak{C}}

\newcommand\ttB         {\mathtt{B}}

\usepackage{mathtools,amssymb,scalerel}

\newcommand\biopencrossl{%
	\mathrel{\scalerel*{>\kern-.4\LMpt\joinrel\blacktriangleleft}{x}}}
\newcommand\biopencrossr{%
	\mathrel{\scalerel*{\blacktriangleright\joinrel\kern-.4\LMpt<}{x}}}



%% file: Jiabib.bib
@Article{Caratheodory1911,
author={Carath{\'e}odory, C.},
title={{\"U}ber den variabilit{\"a}tsbereich der fourier'schen konstanten von positiven harmonischen funktionen},
journal={Rendiconti del Circolo Matematico di Palermo (1884-1940)},
year={1911},
month={Dec},
day={01},
volume={32},
number={1},
pages={193-217},
issn={0009-725X},
doi={10.1007/BF03014795},
url={https://doi.org/10.1007/BF03014795}
}

@article{kang2026quantum,
  title={Quantum Monge-Kantorovich Optimal Transport over Time},
  author={Kang, Siyu and Xiao, Mei-Hui and Peng, Cheng and Jia, Zhian},
  journal={In preparation},
  year={2024}
}

@article{sion1958general,
  title={On general minimax theorems.},
  author={Sion, Maurice},
   journal  = {Pacific Journal of Mathematics},
  year     = {1958},
  volume   = {8},
  number   = {1},
  pages    = {171--176},
  doi      = {10.2140/pjm.1958.8.171},
  mrnumber = {0097026},
  zbl      = {0081.11502}
}

@article{Shi2024Coherence,
  title = {Coherence quantifier based on the quantum optimal transport cost},
  author = {Shi, Xian},
  journal = {Phys. Rev. A},
  volume = {109},
  issue = {5},
  pages = {052443},
  numpages = {7},
  year = {2024},
  month = {May},
  publisher = {American Physical Society},
  doi = {10.1103/PhysRevA.109.052443},
  url = {https://link.aps.org/doi/10.1103/PhysRevA.109.052443}
}

@misc{hooft2009dimensionalreductionquantumgravity,
      title={Dimensional Reduction in Quantum Gravity}, 
      author={G. 't Hooft},
      year={2009},
      eprint={gr-qc/9310026},
      archivePrefix={arXiv},
      primaryClass={gr-qc},
      url={https://arxiv.org/abs/gr-qc/9310026}, 
}

@article{diaz2024STQMBosonFermi,
  title = {Spacetime quantum mechanics for bosonic and fermionic systems},
  author = {Diaz, N. L. and Rossignoli, R.},
  journal = {Phys. Rev. Res.},
  volume = {8},
  issue = {1},
  pages = {013132},
  numpages = {31},
  year = {2026},
  month = {Feb},
  publisher = {American Physical Society},
  doi = {10.1103/zmky-mmwb},
  url = {https://link.aps.org/doi/10.1103/zmky-mmwb}
}

@article{diaz2024STQM,
  title = {Spacetime quantum and classical mechanics with dynamical foliation},
  author = {Diaz, N. L. and Matera, J. M. and Rossignoli, R.},
  journal = {Phys. Rev. D},
  volume = {109},
  issue = {10},
  pages = {105008},
  numpages = {28},
  year = {2024},
  month = {May},
  publisher = {American Physical Society},
  doi = {10.1103/PhysRevD.109.105008},
  url = {https://link.aps.org/doi/10.1103/PhysRevD.109.105008}
}

@misc{hashimoto2026holographyoptimaltransportemergent,
      title={Holography and Optimal Transport: Emergent Wasserstein Spacetime in Harmonic Oscillator, SYK and Krylov Complexity}, 
      author={Koji Hashimoto and Norihiro Tanahashi},
      year={2026},
      eprint={2604.17649},
      archivePrefix={arXiv},
      primaryClass={hep-th},
      url={https://arxiv.org/abs/2604.17649}, 
}

@misc{hashimoto2026wassersteinspacequantumchaos,
      title={Wasserstein Space of Quantum Chaos}, 
      author={Koji Hashimoto and Norihiro Tanahashi and Kentaroh Yoshida},
      year={2026},
      eprint={2605.20995},
      archivePrefix={arXiv},
      primaryClass={hep-th},
      url={https://arxiv.org/abs/2605.20995}, 
}

@article{watrous2009semidefinite,
  title={Semidefinite programs for completely bounded norms},
  author={Watrous, John},
  journal={Theory of Computing},
  volume={5},
  number={11},
  year={2009}
}

@article{watrous2012simpler,
  title={Simpler semidefinite programs for completely bounded norms},
  author={Watrous, John},
  journal={Chicago Journal of Theoretical Computer Science},
  year={2013}
}

@article{vandenberghe1996semidefinite,
  title={Semidefinite programming},
  author={Vandenberghe, Lieven and Boyd, Stephen},
  journal={SIAM review},
  volume={38},
  number={1},
  pages={49--95},
  year={1996},
  publisher={SIAM}
}

@article{Zhou2022OPT,
    author = {Zhou, Li and Yu, Nengkun and Ying, Shenggang and Ying, Mingsheng},
    title = {Quantum earth mover’s distance, a no-go quantum Kantorovich–Rubinstein theorem, and quantum marginal problem},
    journal = {Journal of Mathematical Physics},
    volume = {63},
    number = {10},
    pages = {102201},
    year = {2022},
    month = {10},
    issn = {0022-2488},
    doi = {10.1063/5.0068344},
    url = {https://doi.org/10.1063/5.0068344},
}

@article{De2023Wasserstein,
	author = {{De Palma}, Giacomo and Trevisan, Dario},
	title = {The Wasserstein Distance of Order 1 for Quantum Spin Systems on Infinite Lattices},
	journal = {Annales Henri Poincaré},
	year = {2023},
	month = {12},
	volume = {24},
	number = {12},
	pages = {4237-4282},
	doi = {10.1007/s00023-023-01340-y},
	url = {https://doi.org/10.1007/s00023-023-01340-y},
}

@ARTICLE{DePalma2021Wasserstein,
  author={De Palma, Giacomo and Marvian, Milad and Trevisan, Dario and Lloyd, Seth},
  journal={IEEE Transactions on Information Theory}, 
  title={The Quantum Wasserstein Distance of Order 1}, 
  year={2021},
  volume={67},
  number={10},
  pages={6627-6643},
  doi={10.1109/TIT.2021.3076442}}

@article{Kiani2022OPT,
doi = {10.1088/2058-9565/ac79c9},
url = {https://doi.org/10.1088/2058-9565/ac79c9},
year = {2022},
month = {jul},
publisher = {IOP Publishing},
volume = {7},
number = {4},
pages = {045002},
author = {Kiani, Bobak Toussi and De Palma, Giacomo and Marvian, Milad and Liu, Zi-Wen and Lloyd, Seth},
title = {Learning quantum data with the quantum earth mover’s distance},
journal = {Quantum Science and Technology},
}

@article{DePalma2023OPT,
  title = {Limitations of Variational Quantum Algorithms: A Quantum Optimal Transport Approach},
  author = {De Palma, Giacomo and Marvian, Milad and Rouz{\'e}, Cambyse and Fran{\ifmmode \mbox{\c{c}}\else \c{c}\fi{}}a, Daniel Stilck},
  journal = {PRX Quantum},
  volume = {4},
  issue = {1},
  pages = {010309},
  numpages = {30},
  year = {2023},
  month = {Jan},
  publisher = {American Physical Society},
  doi = {10.1103/PRXQuantum.4.010309},
  url = {https://link.aps.org/doi/10.1103/PRXQuantum.4.010309}
}

@article{vaserstein1969markov,
  title={Markov processes over denumerable products of spaces, describing large systems of automata},
  author={Vaserstein, Leonid Nisonovich},
  journal={Problemy Peredachi Informatsii},
  volume={5},
  number={3},
  pages={64--72},
  year={1969},
}

@book{villani2009optimal,
  title={Optimal transport: old and new},
  author={Villani, C{\'e}dric},
  volume={338},
  year={2009},
  publisher={Springer}
}

@misc{hoogstederriera2026PDOMK,
      title={Approach to optimal quantum transport via states over time}, 
      author={Matt Hoogsteder-Riera and John Calsamiglia and Andreas Winter},
      year={2026},
      eprint={2504.04856},
      archivePrefix={arXiv},
      primaryClass={quant-ph},
      url={https://arxiv.org/abs/2504.04856}, 
}

@article{DePalma2021QuantumMKchannal,
  author = {De Palma, Giacomo and Trevisan, Dario},
  title = {Quantum Optimal Transport with Quantum Channels},
  journal = {Annales Henri Poincaré},
  year = {2021},
  volume = {22},
  number = {10},
  pages = {3199--3234},
  doi = {10.1007/s00023-021-01042-3},
  url = {https://doi.org/10.1007/s00023-021-01042-3},
  isbn = {1424-0661},
   eprint={1911.00803},
  archivePrefix={arXiv},
  primaryClass={quant-ph}  
}

@article{kantorovich2006translocation,
  title={On the Translocation of Masses.},
  author={Kantorovich, Leonid V},
  journal={Journal of mathematical sciences},
  volume={133},
  number={4},
  year={2006},
  url={https://link.springer.com/article/10.1007/s10958-006-0049-2}
}

@article{monge1781memoire,
  title={M{\'e}moire sur la th{\'e}orie des d{\'e}blais et des remblais},
  author={Monge, Gaspard},
  journal={Mem. Math. Phys. Acad. Royale Sci.},
  pages={666--704},
  year={1781}
}

@article{zyczkowski2001monge,
doi = {10.1088/0305-4470/34/34/311},
url = {https://doi.org/10.1088/0305-4470/34/34/311},
year = {2001},
month = {aug},
publisher = {},
volume = {34},
number = {34},
pages = {6689},
author = {Karol Zyczkowski and Wojciech Slomczynski},
title = {The Monge metric on the sphere  and geometry of quantum
states},
journal = {Journal of Physics A: Mathematical and General},
}

@article{zyczkowski1998monge,
doi = {10.1088/0305-4470/31/45/009},
url = {https://doi.org/10.1088/0305-4470/31/45/009},
year = {1998},
month = {nov},
publisher = {},
volume = {31},
number = {45},
pages = {9095},
author = {Karol Zyczkowski and Wojeciech Slomczynski},
title = {The Monge distance between quantum states},
journal = {Journal of Physics A: Mathematical and General},
}

@article{chakrabarti2019quantum,
  title={Quantum Wasserstein generative adversarial networks},
  author={Chakrabarti, Shouvanik and Yiming, Huang and Li, Tongyang and Feizi, Soheil and Wu, Xiaodi},
  journal={Advances in Neural Information Processing Systems},
  volume={32},
  year={2019}
}

@article{Caglioti2023Towards,
  author  = {Caglioti, Emanuele and Golse, Fran{\c{c}}ois and Paul, Thierry},
  title   = {Towards optimal transport for quantum densities},
  journal = {Annali della Scuola Normale Superiore di Pisa, Classe di Scienze},
  year    = {2023},
  volume  = {24},
  number  = {4},
  pages   = {1981--2045},
  doi     = {10.2422/2036-2145.202106_011},
  url     = {https://doi.org/10.2422/2036-2145.202106_011}
}

@article{Caputo2025OPT,
title = {Quantum optimal transport with convex regularization},
journal = {Journal of Functional Analysis},
volume = {290},
number = {4},
pages = {111262},
year = {2026},
issn = {0022-1236},
doi = {https://doi.org/10.1016/j.jfa.2025.111262},
url = {https://www.sciencedirect.com/science/article/pii/S0022123625004446},
author = {Emanuele Caputo and Augusto Gerolin and Nataliia Monina and Lorenzo Portinale},
}

@article{Toth2023quantumwasserstein,
  doi = {10.22331/q-2023-10-16-1143},
  url = {https://doi.org/10.22331/q-2023-10-16-1143},
  title = {Quantum {W}asserstein distance based on an optimization over separable states},
  author = {T{\'{o}}th, G{\'{e}}za and Pitrik, J{\'{o}}zsef},
  journal = {{Quantum}},
  issn = {2521-327X},
  publisher = {{Verein zur F{\"{o}}rderung des Open Access Publizierens in den Quantenwissenschaften}},
  volume = {7},
  pages = {1143},
  month = oct,
  year = {2023}
}

@article{Caglioti2020Quantum,
	author = {Caglioti, E. and Golse, F. and Paul, T.},
	title = {Quantum Optimal Transport is Cheaper},
	journal = {Journal of Statistical Physics},
	year = {2020},
	month = {10},
	volume = {181},
	number = {1},
	pages = {149-162},
	doi = {10.1007/s10955-020-02571-7},
	url = {https://doi.org/10.1007/s10955-020-02571-7},
}

@article{Cole2023On,
	author = {Cole, Sam and Eckstein, Michał and Friedland, Shmuel and Życzkowski, Karol},
	title = {On Quantum Optimal Transport},
	journal = {Mathematical Physics, Analysis and Geometry},
	year = {2023},
	month = {6},
	volume = {26},
	number = {2},
	pages = {14},
	doi = {10.1007/s11040-023-09456-7},
	url = {https://doi.org/10.1007/s11040-023-09456-7},
}

@article{Friedland2020quantumOP,
  title = {Quantum Monge-Kantorovich Problem and Transport Distance between Density Matrices},
  author = {Friedland, Shmuel and Eckstein, Micha{\l{}} and Cole, Sam and {\ifmmode \dot{Z}\else \.{Z}\fi{}}yczkowski, Karol},
  journal = {Phys. Rev. Lett.},
  volume = {129},
  issue = {11},
  pages = {110402},
  numpages = {7},
  year = {2022},
  month = {Sep},
  publisher = {American Physical Society},
  doi = {10.1103/PhysRevLett.129.110402},
  url = {https://link.aps.org/doi/10.1103/PhysRevLett.129.110402}
}

@article{Winter2016Tight,
	author = {Winter, Andreas},
	title = {Tight Uniform Continuity Bounds for Quantum Entropies: Conditional Entropy, Relative Entropy Distance and Energy Constraints},
	journal = {Communications in Mathematical Physics},
	year = {2016},
	month = {10},
	volume = {347},
	number = {1},
	pages = {291-313},
	doi = {10.1007/s00220-016-2609-8},
	url = {https://doi.org/10.1007/s00220-016-2609-8},
}

@article{Golse2016On,
	author = {Golse, François and Mouhot, Clément and Paul, Thierry},
	title = {On the Mean Field and Classical Limits of Quantum Mechanics},
	journal = {Communications in Mathematical Physics},
	year = {2016},
	month = {4},
	volume = {343},
	number = {1},
	pages = {165-205},
	doi = {10.1007/s00220-015-2485-7},
	url = {https://doi.org/10.1007/s00220-015-2485-7},
}

@article{debivre2026kirkwooddirac,
  title = {The Kirkwood-Dirac Representation Associated to the Fourier Transform for Finite Abelian Groups: Positivity},
  author = {De Bièvre, Stephan and Langrenez, Christopher and Radchenko, Danylo},
  year = {2026},
  journal = {Annales Henri Poincaré},
  volume = {27},
  number = {8},
  pages = {3093--3112},
  doi = {10.1007/s00023-025-01614-7},
  url = {https://doi.org/10.1007/s00023-025-01614-7},
  isbn = {1424-0661},
}

@article{DeBievre2021KD,
  title = {Complete Incompatibility, Support Uncertainty, and Kirkwood-Dirac Nonclassicality},
  author = {De Bi{\`e}vre, Stephan},
  journal = {Phys. Rev. Lett.},
  volume = {127},
  issue = {19},
  pages = {190404},
  numpages = {6},
  year = {2021},
  month = {Nov},
  publisher = {American Physical Society},
  doi = {10.1103/PhysRevLett.127.190404},
  url = {https://link.aps.org/doi/10.1103/PhysRevLett.127.190404}
}

@article{Duffin1952frame,
 ISSN = {00029947, 10886850},
 URL = {http://www.jstor.org/stable/1990760},
 author = {R. J. Duffin and A. C. Schaeffer},
 journal = {Transactions of the American Mathematical Society},
 number = {2},
 pages = {341--366},
 publisher = {American Mathematical Society},
 title = {A Class of Nonharmonic Fourier Series},
 urldate = {2026-10-05},
 volume = {72},
 year = {1952}
}

@article{Ferrie2010negativity,
  title = {Necessity of negativity in quantum theory},
  author = {Ferrie, Christopher and Morris, Ryan and Emerson, Joseph},
  journal = {Phys. Rev. A},
  volume = {82},
  issue = {4},
  pages = {044103},
  numpages = {4},
  year = {2010},
  month = {Oct},
  publisher = {American Physical Society},
  doi = {10.1103/PhysRevA.82.044103},
  url = {https://link.aps.org/doi/10.1103/PhysRevA.82.044103}
}

@misc{jia2026unifiedspatiotemporalquantumstates,
      title={Unified spatiotemporal quantum states and spatiotemporal entanglement from Kirkwood-Dirac phase space}, 
      author={Zhian Jia and Mei-Hui Xiao},
      year={2026},
      eprint={2609.24680},
      archivePrefix={arXiv},
      primaryClass={quant-ph},
      url={https://arxiv.org/abs/2609.24680}, 
}

@article{Guo2026Spacetime,
  author = {Guo, Wu-zhong},
  title = {Spacetime density matrix: formalism and properties},
  journal = {Journal of High Energy Physics},
  year = {2026},
  volume = {2026},
  number = {1},
  pages = {128},
  doi = {10.1007/JHEP01(2026)128},
  url = {https://doi.org/10.1007/JHEP01(2026)128},
  isbn = {1029-8479}
}

@misc{diaz2026unifyingspacetimeapproachesquantum,
      title={Unifying spacetime approaches to quantum mechanics}, 
      author={N. L. Diaz and M. Cerezo and Paolo Braccia},
      year={2026},
      eprint={2606.12539},
      archivePrefix={arXiv},
      primaryClass={quant-ph},
      url={https://arxiv.org/abs/2606.12539}, 
}

@article{Rovelli1991time,
  title = {Time in quantum gravity: An hypothesis},
  author = {Rovelli, Carlo},
  journal = {Phys. Rev. D},
  volume = {43},
  issue = {2},
  pages = {442--456},
  numpages = {0},
  year = {1991},
  month = {Jan},
  publisher = {American Physical Society},
  doi = {10.1103/PhysRevD.43.442},
  url = {https://link.aps.org/doi/10.1103/PhysRevD.43.442}
}

@article{Page1983evolution,
  title = {Evolution without evolution: Dynamics described by stationary observables},
  author = {Page, Don N. and Wootters, William K.},
  journal = {Phys. Rev. D},
  volume = {27},
  issue = {12},
  pages = {2885--2892},
  numpages = {0},
  year = {1983},
  month = {Jun},
  publisher = {American Physical Society},
  doi = {10.1103/PhysRevD.27.2885},
  url = {https://link.aps.org/doi/10.1103/PhysRevD.27.2885}
}

@article{Watanabe1955twostateQM,
  title = {Symmetry of Physical Laws. Part III. Prediction and Retrodiction},
  author = {Watanabe, Satosi},
  journal = {Rev. Mod. Phys.},
  volume = {27},
  issue = {2},
  pages = {179--186},
  numpages = {0},
  year = {1955},
  month = {Apr},
  publisher = {American Physical Society},
  doi = {10.1103/RevModPhys.27.179},
  url = {https://link.aps.org/doi/10.1103/RevModPhys.27.179}
}

@misc{aharonov2007twostatevectorformalismqauntum,
      title={The Two-State Vector Formalism of Qauntum Mechanics: an Updated Review}, 
      author={Yakir Aharonov and Lev Vaidman},
      year={2007},
      eprint={quant-ph/0105101},
      archivePrefix={arXiv},
      primaryClass={quant-ph},
      url={https://arxiv.org/abs/quant-ph/0105101}, 
}

@article{Lerose2021influencematrix,
  title = {Influence Matrix Approach to Many-Body Floquet Dynamics},
  author = {Lerose, Alessio and Sonner, Michael and Abanin, Dmitry A.},
  journal = {Phys. Rev. X},
  volume = {11},
  issue = {2},
  pages = {021040},
  numpages = {21},
  year = {2021},
  month = {May},
  publisher = {American Physical Society},
  doi = {10.1103/PhysRevX.11.021040},
  url = {https://link.aps.org/doi/10.1103/PhysRevX.11.021040}
}

@article{Feynman1963influenceMatrix,
title = {The theory of a general quantum system interacting with a linear dissipative system},
journal = {Annals of Physics},
volume = {24},
pages = {118-173},
year = {1963},
issn = {0003-4916},
doi = {https://doi.org/10.1016/0003-4916(63)90068-X},
url = {https://www.sciencedirect.com/science/article/pii/000349166390068X},
author = {R.P Feynman and F.L Vernon}
}

@misc{milekhin2025observable,
      title={Observable and computable entanglement in time}, 
      author={Alexey Milekhin and Zofia Adamska and John Preskill},
      year={2025},
      eprint={2502.12240},
      archivePrefix={arXiv},
      primaryClass={quant-ph},
      url={https://arxiv.org/abs/2502.12240}, 
}

@Article{Glorioso2024spacetimeMutual,
author={Glorioso, Paolo
and Qi, Xiao-Liang
and Yang, Zhenbin},
title={Space-time generalization of mutual information},
journal={Journal of High Energy Physics},
year={2024},
month={May},
day={31},
volume={2024},
number={5},
pages={338},
issn={1029-8479},
doi={10.1007/JHEP05(2024)338},
url={https://doi.org/10.1007/JHEP05(2024)338},
  eprint={2401.02475},
  archivePrefix={arXiv},
  primaryClass={quant-ph} 
}

@misc{jia2026temporaltomography,
      title={Temporal State Tomography via Quantum Snapshotting the Temporal Quasiprobabilities}, 
      author={Zhian Jia},
      year={2026},
      eprint={2605.02655},
      archivePrefix={arXiv},
      primaryClass={quant-ph},
      url={https://arxiv.org/abs/2605.02655}, 
}

@article{Lie2025stateovertime,
  title = {Multipartite Quantum States over Time from Two Fundamental Assumptions},
  author = {Lie, Seok Hyung and Fullwood, James},
  journal = {Phys. Rev. Lett.},
  volume = {135},
  issue = {23},
  pages = {230204},
  numpages = {8},
  year = {2025},
  month = {Dec},
  publisher = {American Physical Society},
  doi = {10.1103/lbf3-snp8},
  url = {https://link.aps.org/doi/10.1103/lbf3-snp8}
}

@misc{Jia2025TemporalKirkwoodDirac,
   title={Temporal Kirkwood-Dirac Quasiprobability Distribution and Unification of Temporal State Formalisms through Temporal Bloch Tomography}, 
      author={Zhian Jia and Kavan Modi and Dagomir Kaszlikowski},
      year={2026},
      eprint={2601.05294},
      archivePrefix={arXiv},
      primaryClass={quant-ph},
      url={https://arxiv.org/abs/2601.05294}, 
}

@article{margenau1961correlation,
  title={Correlation between measurements in quantum theory},
  author={Margenau, Henry and Hill, Robert Nyden},
  journal={Progress of Theoretical Physics},
  volume={26},
  number={5},
  pages={722--738},
  year={1961},
  publisher={Oxford University Press},
  url={https://academic.oup.com/ptp/article/26/5/722/1936017},
  doi={10.1143/PTP.26.722},
}

@article{jia2024spatiotemporal,
  title={The Spatiotemporal Doubled-Density Operator: A Unified Framework for Analyzing Spatial and Temporal Quantum Processes},
  author={Jia, Zhian and Kaszlikowski, Dagomir},
  journal={Advanced Quantum Technologies},
  volume={7},
  number={11},
  pages={2400102},
  year={2024},
  publisher={Wiley Online Library},
  url={https://advanced.onlinelibrary.wiley.com/doi/10.1002/qute.202400102},
  doi = {10.1002/qute.202400102},
  eprint={2305.15649},
  archivePrefix={arXiv},
  primaryClass={quant-ph}   
}

@article{Chiribella2009comb,
  title = {Theoretical framework for quantum networks},
  author = {Chiribella, Giulio and D'Ariano, Giacomo Mauro and Perinotti, Paolo},
  journal = {Phys. Rev. A},
  volume = {80},
  issue = {2},
  pages = {022339},
  numpages = {20},
  year = {2009},
  month = {Aug},
  publisher = {American Physical Society},
  doi = {10.1103/PhysRevA.80.022339},
  url = {https://link.aps.org/doi/10.1103/PhysRevA.80.022339},
  eprint={0904.4483},
  archivePrefix={arXiv},
  primaryClass={quant-ph}   
}

@article{Aharonov2009multi,
  title = {Multiple-time states and multiple-time measurements in quantum mechanics},
  author = {Aharonov, Yakir and Popescu, Sandu and Tollaksen, Jeff and Vaidman, Lev},
  journal = {Phys. Rev. A},
  volume = {79},
  issue = {5},
  pages = {052110},
  numpages = {16},
  year = {2009},
  month = {May},
  publisher = {American Physical Society},
  doi = {10.1103/PhysRevA.79.052110},
  url = {https://link.aps.org/doi/10.1103/PhysRevA.79.052110},
  eprint={0712.0320},
  archivePrefix={arXiv},
  primaryClass={quant-ph}    
}

@article{Parzygnat2023pdo,
  title = {From Time-Reversal Symmetry to Quantum Bayes' Rules},
  author = {Parzygnat, Arthur J. and Fullwood, James},
  journal = {PRX Quantum},
  volume = {4},
  issue = {2},
  pages = {020334},
  numpages = {26},
  year = {2023},
  month = {Jun},
  publisher = {American Physical Society},
  doi = {10.1103/PRXQuantum.4.020334},
  url = {https://link.aps.org/doi/10.1103/PRXQuantum.4.020334},
eprint={2212.08088},
  archivePrefix={arXiv},
  primaryClass={quant-ph}  
}

@article {liu2023unification,
    AUTHOR = {Liu, Xiangjing and Jia, Zhian and Qiu, Yixian and Li, Fei and
              Dahlsten, Oscar},
     TITLE = {Unification of spatiotemporal quantum formalisms: mapping
              between process and pseudo-density matrices via multiple-time
              states},
   JOURNAL = {New J. Phys.},
  FJOURNAL = {New Journal of Physics},
    VOLUME = {26},
      YEAR = {2024},
    NUMBER = {March},
     PAGES = {Paper No. 033008, 15},
   MRCLASS = {81P16},
  MRNUMBER = {4730855},
       DOI = {10.1088/1367-2630/ad264c},
       URL = {https://doi.org/10.1088/1367-2630/ad264c},
    eprint={2306.05958},
  archivePrefix={arXiv},
  primaryClass={quant-ph}   
}

@article{fullwood2022quantum,
  title={On quantum states over time},
  author={Fullwood, James and Parzygnat, Arthur J},
  journal={Proceedings of the Royal Society A},
  volume={478},
  number={2264},
  pages={20220104},
  year={2022},
  publisher={The Royal Society},
  url={https://royalsocietypublishing.org/doi/abs/10.1098/rspa.2022.0104},
    eprint={2202.03607},
  archivePrefix={arXiv},
  primaryClass={quant-ph}   
}

@article{Dirac1945on,
  title = {On the Analogy Between Classical and Quantum Mechanics},
  author = {Dirac, P. A. M.},
  journal = {Rev. Mod. Phys.},
  volume = {17},
  issue = {2-3},
  pages = {195--199},
  numpages = {0},
  year = {1945},
  month = {Apr},
  publisher = {American Physical Society},
  doi = {10.1103/RevModPhys.17.195},
  url = {https://link.aps.org/doi/10.1103/RevModPhys.17.195}
}

@article{Kirkwood1933quantum,
  title = {Quantum Statistics of Almost Classical Assemblies},
  author = {Kirkwood, John G.},
  journal = {Phys. Rev.},
  volume = {44},
  issue = {1},
  pages = {31--37},
  numpages = {0},
  year = {1933},
  month = {Jul},
  publisher = {American Physical Society},
  doi = {10.1103/PhysRev.44.31},
  url = {https://link.aps.org/doi/10.1103/PhysRev.44.31}
}

@article {jia2023quantumspace,
    AUTHOR = {Jia, Zhian and Song, Minjeong and Kaszlikowski, Dagomir},
     TITLE = {Quantum space-time marginal problem: global causal structure
              from local causal information},
   JOURNAL = {New J. Phys.},
  FJOURNAL = {New Journal of Physics},
    VOLUME = {25},
      YEAR = {2023},
    NUMBER = {December},
     PAGES = {Paper No. 123038, 23},
   MRCLASS = {81P45 (81P16 81P40)},
  MRNUMBER = {4687136},
       DOI = {10.1088/1367-2630/ad1416},
       URL = {https://doi.org/10.1088/1367-2630/ad1416},
  eprint={2303.12819},
  archivePrefix={arXiv},
  primaryClass={quant-ph} 
}

@inproceedings{gutoski2007toward,
 title={Toward a general theory of quantum games},
  author={Gutoski, Gus and Watrous, John},
  booktitle={Proceedings of the thirty-ninth annual ACM symposium on Theory of computing},
  pages={565--574},
  year={2007},
  url={https://dl.acm.org/doi/10.1145/1250790.1250873},
  eprint={quant-ph/0611234},
  archivePrefix={arXiv},
  primaryClass={quant-ph}    
}

@article{cotler2018superdensity,
  title={Superdensity operators for spacetime quantum mechanics},
  author={Cotler, Jordan and Jian, Chao-Ming and Qi, Xiao-Liang and Wilczek, Frank},
  journal={Journal of High Energy Physics},
  volume={2018},
  number={9},
  pages={1--57},
  year={2018},
  publisher={Springer},
  url={https://link.springer.com/article/10.1007/JHEP09(2018)093},
  eprint={1711.03119 },
  archivePrefix={arXiv},
  primaryClass={quant-ph}       
}

@article{griffiths1984consistent,
  title={Consistent histories and the interpretation of quantum mechanics},
  author={Griffiths, Robert B},
  journal={Journal of Statistical Physics},
  volume={36},
  number={1},
  pages={219--272},
  year={1984},
  publisher={Springer},
  url={https://link.springer.com/article/10.1007/BF01015734}
}

@article{oreshkov2012quantum,
  title={Quantum correlations with no causal order},
  author={Oreshkov, Ognyan and Costa, Fabio and Brukner, {\v{C}}aslav},
  journal={Nature Communications},
  volume={3},
  number={1},
  pages={1--8},
  year={2012},
  publisher={Nature Publishing Group},
  url={https://www.nature.com/articles/ncomms2076},
  eprint={1105.4464},
  archivePrefix={arXiv},
  primaryClass={quant-ph}     
}

@article{ferrie2011quasi,
  title={Quasi-probability representations of quantum theory with applications to quantum information science},
  author={Ferrie, Christopher},
  journal={Reports on Progress in Physics},
  volume={74},
  number={11},
  pages={116001},
  year={2011},
  publisher={IOP Publishing},
  url={https://iopscience.iop.org/article/10.1088/0034-4885/74/11/116001},
  eprint={1010.2701},
  archivePrefix={arXiv},
  primaryClass={quant-ph} 
}

@article{fitzsimons2015quantum,
  title={Quantum correlations which imply causation},
  author={Fitzsimons, Joseph F and Jones, Jonathan A and Vedral, Vlatko},
  journal={Scientific Reports},
  volume={5},
  number={1},
  pages={1--7},
  year={2015},
  publisher={Nature Publishing Group},
  url={https://www.nature.com/articles/srep18281},
  eprint={1302.2731},
  archivePrefix={arXiv},
  primaryClass={quant-ph} 
}

@Article{Maldacena1999,
  author   = {Maldacena, Juan},
  title    = {The Large-N Limit of Superconformal Field Theories and Supergravity},
  journal  = {International Journal of Theoretical Physics},
  year     = {1999},
  volume   = {38},
  number   = {4},
  pages    = {1113--1133},
  month    = {Apr},
  issn     = {1572-9575},
  day      = {01},
  doi      = {10.1023/A:1026654312961},
  url      = {https://doi.org/10.1023/A:1026654312961},
}

@Article{witten1998anti,
  author  = {Witten, E},
  title   = {Anti-de Sitter space, thermal phase transition and confinement in gauge theories},
  journal = {Adv. Theor. Math. Phys},
  year    = {1998},
  volume  = {2},
  pages   = {505},
}

@Article{Susskind1995,
  author  = {Susskind,Leonard},
  title   = {The world as a hologram},
  journal = {Journal of Mathematical Physics},
  year    = {1995},
  volume  = {36},
  number  = {11},
  pages   = {6377-6396},
  doi     = {10.1063/1.531249},
  url     = { https://doi.org/10.1063/1.531249},
}

@article{Pollock2018processtensor,
  title = {Non-Markovian quantum processes: Complete framework and efficient characterization},
  author = {Pollock, Felix A. and Rodr\'{\i}guez-Rosario, C\'esar and Frauenheim, Thomas and Paternostro, Mauro and Modi, Kavan},
  journal = {Phys. Rev. A},
  volume = {97},
  issue = {1},
  pages = {012127},
  numpages = {13},
  year = {2018},
  month = {Jan},
  publisher = {American Physical Society},
  doi = {10.1103/PhysRevA.97.012127},
  url = {https://link.aps.org/doi/10.1103/PhysRevA.97.012127}
}
